\documentclass[apjl]{emulateapj}
\usepackage{apjfonts}

\def\plotfiddle#1#2#3#4#5#6#7{\centering \leavevmode
    \vbox to#2{\rule{0pt}{#2}}
    \includegraphics{#1}}

\def\alwaysmath#1{\ifmmode{#1}\else{$#1$}\fi}

\lefthead{Mu\~noz et al.}
\righthead{Modeling Dwarf Spheroidals with Dark Matter and Tides}

\begin{document}
\title{Modeling the Structure and Dynamics of Dwarf Spheroidal Galaxies 
with Dark Matter and Tides}

\author{
Ricardo R. Mu\~noz\altaffilmark{1},
Steven R. Majewski\altaffilmark{1}, 
Kathryn V. Johnston\altaffilmark{2,3} 
}

\altaffiltext{1}{Dept. of Astronomy, University of Virginia,
Charlottesville, VA 22904-4325 (rrm8f, srm4n@virginia.edu)}

\altaffiltext{2}{Department of Astronomy, Wesleyan University, Middletown, CT}

\altaffiltext{3}{Currently in the Department of Astronomy at Columbia University, New York City, NY
(kvj@astro.columbia.edu)}

\begin{abstract}
Two observed features of dwarf spheroidal (dSph) galaxies --- extended 
light profiles and more or less flat velocity dispersion profiles --- have led 
to competing interpretations over the nature of dSphs: Either the 
outer characteristics of dSphs are reflecting ongoing tidal disruption, or 
they are showing the existence of bound stellar ``halos" embedded within 
extended dark matter halos. To accommodate the observed dSph features 
within the latter interpretation, dSph models now commonly adopt multi-component
structures with strongly varying $M/L$ with radius.  
In contrast, we explore whether the observed features 
of dSphs can still be accounted for with simpler, mass-follows-light 
(MFL) models including tidal disruption.  As a test case, we focus on 
modeling the Carina dSph, which resembles other classical dSphs in shape and
velocity distribution but presently has the most extensive data at large radius.  
We find that previous $N$-body, MFL simulations of (tidally disrupting) dSphs  
did not sufficiently explore the parameter space of satellite mass, density and 
orbital shape to find adequate matches to Galactic dSph systems, 
whereas with a systematic survey of parameter space we are able to find
tidally disrupting, MFL satellite 
models that rather faithfully reproduce Carina's velocity profile, velocity 
dispersion profile and projected density distribution over its entire sampled radius. 
The successful MFL model satellites have very eccentric orbits, currently favored 
by Cold Dark Matter (CDM) models, and central velocity dispersions that still yield
an accurate representation of the bound mass and observed central $M/L\sim40$ of 
Carina, despite inflation of the velocity dispersion outside the dSph core by unbound 
debris.  Our survey of parameter space also allows us to address a number of commonly
held misperceptions of tidal disruption and its observable effects on dSph structure 
and dynamics.  The simulations suggest that (1) even modest tidal disruption can have 
a profound effect on the observed dynamics of dSph stars at large radii, and (2) it may 
be premature to abandon single-component, MFL models, which can successfully reproduce 
all salient structural and dynamical properties of the Carina (and other) dSph satellites 
without invoking special timing or viewing perspectives.  Satellites that are well-described 
by tidally disrupting MFL models could still be fully compatible with $\Lambda$-CDM 
if for example they represent a later stage in the evolution of luminous subhalos.
\end{abstract}


\section{Introduction}

\subsection{Dark Matter and Tidal Effects in dSphs: Previous Work}

Ever since the work of Hodge and Michie \citep{Hodge1961a,Hodge1961b,Hodge1962,HM1969}, which 
suggested that Galactic tides might play an important
role in shaping the structure of Galactic dwarf spheroidal galaxies (dSphs), the extent of 
this tidal influence has been debated.  When   
early precision radial velocity measurements were obtained
in the central regions of a number of these dSphs (\citealt{Aaronson1983};
\citealt{Mateo1993}; \citealt{Hargreaves1994}) the relative importance of tidal forces 
became less clear because the measured high velocity dispersions
implied large dark matter contents within dSphs.
Using traditional core-fitting methods (e.g., \citealt{Illingworth76}; \citealt{Richstone1986}), 
central mass-to-light ratio ($M/L$) values ranging from 
$\sim5$ for Leo I (\citealt{Mateo1998b}; \citealt{Sohn2007}) 
to $\sim100$ for Draco and Ursa Minor (\citealt{Armandroff1995}), 
and even as high as 500 for Ursa Major (\citealt{Kleyna2005}) and 
Bo\"{o}tes (\citealt{Mu06b}; \citealt{Martin2007})
have been found among Galactic dSph galaxies.

Such large $M/L$'s have been difficult to reconcile with
the notion that these dwarfs may be tidally 
truncated (\citealt{Cuddeford1990}; \citealt{Burkert1997}).
Moreover, were they to have such high $M/L$, dSphs would be structurally very
different from systems of relatively similar mass and luminosity scale,
such as globular clusters and dwarf elliptical galaxies.
The notion that dSph velocity dispersions may not directly reflect 
their mass but may be inflated by various mechanisms has prompted frequent assessment of 
various alternative dSph structural models --- both Newtonian and non-Newtonian (e.g., MOND;
\citealt{Milgrom1995}, \citealt{Lokas2001},
but see also, e.g., \citealt{Gerhard1992}, \citealt{Sanchez2007}) --- as well as
the assumptions built into
the core-fitting method for deriving $M/L$'s.\footnote{Core-fitting assumes that these dSphs are in
virial equilibrium, their velocity distributions are isotropic, their mass distribution
follows that of the light, and their surface brightness, or equivalently, their surface
density, are well represented by a \citet{King1962} model.}

For example, among the suggestions within a Newtonian framework,
the possible influence of binary stars to inflate velocity dispersions 
was investigated, but found to be minor (\citealt{Olszewski1996}; \citealt{Hargreaves1996};
\citealt{Kleyna2002}).  
The influence of velocity jitter in the atmospheres of giant stars is also likely 
insignificant since such variations rarely exceed 0.5-1 km s$^{-1}$ (\citealt{Bizyaev2006}) or perhaps 
slightly more \citep{Carney03}.
Other explanations for the seemingly high observed dSph velocity 
dispersions question the validity
of the virial equilibrium assumption, in some cases exploring violations of that 
equilibrium in the extreme
--- i.e. proposed models whereby the dSphs are in complete or
near complete tidal disruption and their appearance of structural cohesion is
an artefact of other dynamical processes that obviate the need for dark 
matter. 
Synchronicity between internal stellar and external bulk dSph orbits, for
instance, has been proposed 
as one means to explain the high velocity dispersions and 
apparent spatial coherence of dSph systems
in the resonant orbit coupling models of \citet{KM1989}
and \citet{Kuhn1993}.
Subsequent work by \citet{Pryor96} and \citet{Sellwood1998} showed that the 
tidal oscillation mechanism proposed by \citet{KM1989} 
is not excited by motion through a realistic, logarithmic Galactic potential,
and that the tidally disrupted remnants proposed by \citet{Kuhn1993} would be
some ten times larger than the observed dSphs. However, \citet{fleck2003}
have updated the \citet{KM1989} model and suggest that tidally induced oscillations 
{\it can} inflate the velocity dispersion by up to an order of magnitude, although only
for a short fraction of a dSph orbital period.

Meanwhile, models of dSphs that are mostly bound but that incorporate the presence of some 
unbound stars have been partly motivated by the discovery that most of the Galactic dSphs have
an extended component beyond their nominal King limiting radius $r_{lim}$ (obtained by fitting a 
King profile  to the inner light distributions; see Appendix A), a feature
suggested to be evidence of tidal disruption in these systems 
(\citealt{Eskridge1988a,Eskridge1988b}; \citealt{IH95}; \citealt{kuhn96}; 
\citealt{Smith1997}; \citealt{PaperII}). 
Since these early photometric studies, extended density profile ``break" 
populations\footnote{Here, ``break" population refers to an empirical description
of the density profile where the outer parts of the density distribution
seems to ``break" to a new, less steep form, departing from
a distribution (usually resembling a King profile --- see Appendix A) 
that fits well the inner parts of Galactic dSphs.} 
have been found and/or verified in the Ursa Minor 
(\citealt{Delgado2001}, \citealt{Palma2003}; \citealt{Mu05}), Sculptor 
(\citealt{Walcher2003}; \citealt{Westfall2006}), Draco (\citealt{W04}), 
Leo I (\citealt{Sohn2007}) and Carina dSphs (\citealt{PaperII,PaperVI}; 
\citealt{Mu06a}).  As has been pointed out in many of these references, the observed ``bimodal" 
dSph structural profiles in these systems 
look very similar to those produced by $N$-body simulations 
of tidally disrupting dwarf galaxies (e.g., \citealt{JZSH99}, \citealt{Mayer2002}; \citealt{Choi2002}), 
and in at least one other real dSph case sharing this radial profile shape --- the 
Sagittarius (Sgr) dSph 
(\citealt{MSWO}) --- the
secondary, extended population is unquestionably constituted by tidally stripped stars.  
By analogy it might seem logical to surmise that the other dSph profiles have been shaped 
by the same disruption processes.  However, two frequently cited $N$-body studies of 
dSph tidal disruption are typically invoked to argue against this conclusion.

In an attempt to investigate the degree to which Galactic tides can affect the structural 
and kinematical properties of dSphs -- and therefore their inferred 
$M/L$ -- \citet[][hereafter {PP95}]{PP95} and \citet[][hereafter {OLA95}]{OLA95} each carried 
out numerical simulations of low density satellites containing no dark matter orbiting in a 
Milky Way (MW)--like potential.  \citeauthor*{OLA95} modeled low mass satellites 
($M_{\rm satellite}<2\times10^{6}$ M$_{\sun}$) orbiting in rather dynamically gentle 
orbits (eccentricities no greater than 0.5, and perigalactica no closer than 50 kpc) having 
several perigalacticon passages.
OLA95 found that for such systems the satellite density profile becomes flattened in the orbital plane
of the galaxy and follows aproximately a power-law distribution with
long strands of tidally stripped debris formed in some cases.
However, \citeauthor*{OLA95} also found that the velocity dispersion of the unbound, but
not yet dispersed, stars was similar to the previous central equilibrium value, implying 
that the observed high velocity dispersions of dSphs were not being significantly inflated 
by tidal stripping; \citeauthor*{OLA95} concluded, therefore, that  substantial 
mass content within the dSphs is unavoidable, even when/if tides are important.

\citeauthor*{PP95} also studied low mass satellites with masses 
of $2.7\times10^5$ M$_{\sun}$ and
$8.1\times10^5$ M$_{\sun}$ but in rather eccentric orbits (apogalactica of 210 kpc
and perigalactica of 30 kpc). Again, rather strongly formed tidal tails were 
formed in these models, but \citeauthor*{PP95} found that a significant, artificial 
(i.e., tidal) increase of the $M/L$ was rarely observed, and, even when found, 
the phenomenon lasts only briefly (several times $10^{8}$ yr); to invoke this scenario 
as an explanation for the high observed central
velocity dispersion in dSphs would require that we live in a very special 
time to be observing all dSphs in the same evolutionary phase (near perigalacticon).   
Thus, \citeauthor*{PP95} 
concluded that it is very unlikely that the central regions of dSphs are 
influenced by tides.
They also pointed out that the effects of Galactic tides on 
dSphs were, for the most part, undetectable, except perhaps for two key features: 
(1) a velocity gradient (apparent rotation) that should be observed along the major axis of tidally 
disrupting dwarf galaxies
and, (2) a surface brightness distribution that should become more elliptical with radius.

The \citeauthor*{OLA95} and \citeauthor*{PP95} studies have been highly 
influential in reinforcing the notion that tidal disruption is {\it not} occuring in 
most dSph systems, since observed dSph galaxies with large central velocity 
dispersions do not resemble {\it these particular} models of disrupted systems, which showed 
tidally induced rotation, distorted morphologies and low central velocity dispersions.
Thus has evolved a frequently presented, ``bimodal" pair of viewpoints that dSph galaxies 
{\it either} have little dark matter and therefore are highly prone to disrupting, {\it or} 
that they have  high dark matter contents that are incommensurate with tidal 
stripping\footnote{Even when dark matter and tides are included in models, the
conclusions lead to a bimodal position. For example, \citet{Read2006b} and \citet{Mashchenko2006} 
combined tidal interaction with dark matter-dominated, two-component satellite models when 
studying Galactic dSphs, but concluded that tides are unimportant due to the immense amounts 
of dark matter in the satellite extended dark matter halos they need to explain the velocity
dispersion profiles of dSphs, effectively reinforcing the ``bimodal" picture of tides versus 
dark matter.} (e.g., \citealt{Burkert1997}; \citealt{Kroupa1997}; \citealt{GFM99}; 
\citealt{Kleyna2001}; \citealt{GFDM03}; \citealt{Gilmore2006}; \citealt{Koch2007a}).
However, though not technically flawed,
the \citeauthor*{OLA95} and \citeauthor*{PP95} studies 
are highly specialized --- the former
pertaining specifically to low mass dSphs on rather circular orbits whereas 
\citeauthor*{PP95} selected  as strawman examples models with $M/L$ so 
unrealistically low that the satellites do not even survive one perigalacticon 
passage ---  so that it is not clear how general are the conclusions of these
studies and whether they should continue to be applied to the
interpretation of the much better data we now have on dSph systems.
For example, in reality it is likely that dSphs have rather eccentric orbits, 
as is now known to be the case for at least several 
of the most recent orbits of the Sgr system (\citealt{Law2005}), as is expected for the 
Leo I system (with its enormous radial velocity), as has been measured for several 
other dSphs (\citealt{Piatek2003,Piatek2005,Piatek2006}) and as is predicted
for cold dark matter (CDM) subhalos (e.g., \citealt{Ghigna1998}).  
Moreover, the PP95-adopted $M/L=1$ $M_{\sun}/L_{\sun}$ is 
several times lower than that expected even for an old stellar population
with no dark matter, so
that the \citeauthor*{PP95} modeled satellites are much fluffier than would be 
expected for any dSph system, even ignoring dark matter.  
While important for demonstrating how dwarf satellites would evolve in 
specific circumstances, these early models should be used cautiously
as guides to understanding real dSph systems on eccentric orbits.

Subsequent to these 1995 studies, \citet{Kroupa1997} and \citet{Klessen1998} were 
successful in explaining many of the salient features observed in present-day dSphs 
by modeling them as long-lived, dark matter free, but mostly {\it unbound} remnant 
dSph cores of mass only a few percent of the initial satellite mass. 
In these simulations, the satellites have physical dimensions comparable to Galactic 
dSphs, they follow the observed linear correlation between surface brightness and 
total luminosity (\citealt{Bellazzini1996}),  and more importantly, they have inferred 
(but artificial) $M/L$ as high as 100.  However, these models only work if the satellites 
are in such an eccentric orbit that we see them and their debris more or less projected 
along the same line of sight. Such alignments are directly testable because the projection 
of tidal debris is expected to result in a significant spread in color magnitude diagram 
(CMD) features like the horizontal branch (\citealt{Klessen2002}). This depth effect should 
be detectable with wide-field photometry, a test carried out by \citet{Klessen2003} with 
public data from the Sloan Digital Sky Survey (SDSS) in the Draco dSph field. They showed 
that Draco's narrow horizontal branch cannot be reproduced by this alignment model and 
concluded that Draco cannot be the unbound remnant of a disrupted galaxy, and must, 
therefore, be dark matter dominated.  This additional ``failure of the tidal 
model"  --- though again pertaining to a prediction from a rather highly specialized 
satellite model --- has further promoted a perception that tidal effects are unimportant 
in dSphs or at least have not affected them significantly (e.g., \citealt{Kleyna2001}; 
\citealt{Walker2006a}; \citealt{Read2006b}; \citealt{Gilmore2006}).

More recently, new, large area radial velocity (RV) datasets have become available for
the Galactic dSphs and allow for the first time the study of the line of sight 
velocity dispersion as a function of radius out to (and in some cases well past)
$r_{lim}$.  Such extensive velocity data provide new and remarkably consistent 
constraints on the structure of dSphs. Flat or only slightly rising/declining 
velocity dispersion profiles have been reported for Sculptor (\citealt{Tolstoy2004}; 
\citealt{Westfall2006}), Draco (\citealt{W04}; \citealt{Mu05}), Ursa Minor (\citealt{W04}; 
\citealt{Mu05}), Fornax (\citealt{Walker2006a}), Sextans (\citealt{Walker2006b}), 
Leo I (\citealt{Sohn2007}; \citealt{Koch2007a}), Leo II (\citealt{Koch2007b}, 
\citealt{Siegel07}), Carina (\citealt{Mu06a}) and Sagittarius (S. R. Majewski et al., 
in preparation).  Previous studies (e.g., \citealt{Kroupa1997}; \citealt{Kleyna1999}) 
have demostrated that an important
difference between a dark matter-dominated, non-disrupting, mass-follows-light (MFL) dSph and
a tidally disrupted one is the radial behavior of the velocity dispersion.
The former model would yield a decreasing dispersion with radius, 
whereas a flat or rising profile is expected for a tidally disrupting dwarf.  
Thus, tidal disruption provides one explanation for the observed dSph dispersion
profiles.

But a third structural possibility, a luminous dSph surrounded by a more 
extended, mostly dark matter halo is also consistent with the new observations.
CDM cosmology-driven models (\citealt{Stoehr2002}; \citealt{Hayashi2003}) 
have shown that the observed dSph velocity dispersion profiles are in good agreement 
with simulations wherein dSphs inhabit the most massive subhalos
(10$^{9}$ -- 10$^{10}$ M$_{\sun}$) in a $\Lambda$-CDM universe, a result that also 
helps mitigate the so-called missing satellite problem inherent to these cosmologies 
(\citealt{Kauffmann1993}; \citealt{Klypin1999}; \citealt{Moore1999}). 
This theoretical result, and the apparent lack of observational evidence supporting
tidal stripping as an ongoing process (in the context of the
\citeauthor*{PP95} and \citeauthor*{OLA95} models), has prompted 
some more recent studies (e.g., \citealt{Kleyna2002};
\citealt{Lokas2002}; \citealt{Mashchenko2006}; \citealt{Walker2006a}; \citealt{Read2006b})
to model the structure of dSphs with two component models (compact stellar and 
extended dark mass components) to account for the observed dSph velocity dispersion 
profiles; these two component models use extended DM halos to keep the velocity
dispersion profiles flat, and then conveniently create total masses consistent with 
the enormous total $M/L$ [exceeding $10^3$ or $10^4$ (M/L)$_{\sun}$] required for dSphs 
to populate the high end of the subhalo mass 
spectrum in $\Lambda$-CDM predictions.\footnote{We note that \citet{W04} reported a sudden
drop in projected velocity dispersion at about the King limiting radius for the Ursa Minor 
and Draco dSphs, and \citet{Kleyna2004} reported a similar finding for Sextans.  
\citet{Read2006b} attempted to model this apparent kinematical feature and concluded 
that (1) dSph galaxies could have masses as high as 10$^{10}$ M$_{\sun}$, and (2) tidal
stripping is unimportant interior to $\sim1$ kpc, although ``tidal shocking" may still be
important.  However, as has been shown by \citet{Lokas2005} and  \citet{Mu05}, the
reported ``cold points" are likely an artifact of binning and 
of ignoring ``velocity distribution outliers". In one case (Ursa Minor) the result 
was not confirmed with additional data \citep{Mu05}.}
These models explicitly assume
that the dwarf satellite is in virial equilibrium thoughout its physical extent, and
therefore, neglect any tidal influence in the shape of the velocity distribution.

\subsection{Goals of this Paper}

Against this backdrop of previous work on dSph structure and tidal influences, the present 
paper has several goals.  
First, we want to readdress the question
to what extent Galactic tides may be affecting dSph galaxies, and,
more specifically, reassess the issue of whether other presently
observed Galactic dSph satellites may be tidally disrupting along the Sgr paradigm.
In some ways our modeling is reminiscent of that by \citeauthor*{OLA95} 
and \citeauthor*{PP95}, but we explore a much wider range of satellite orbit, mass 
and density, and therefore a greater 
breadth of context to test the character of possible resulting tidal disruption (\S2).
We return to
simpler, more prosaic, one-component, MFL satellite structures
because the previous studies have not completely explored the full
parameter space possible within this paradigm
and therefore (in our opinion) have not definitively ruled out MFL models.
It is also possible that MFL dwarfs are a reasonable model for disrupting
two-component models, after the outer envelope of extended dark matter
has been stripped away (\citealt{Klimentowski2007}).

But a key difference between our work and the influential \citeauthor*{OLA95}
and \citeauthor*{PP95} studies is that 
we explore the case of tidal disruption in {\it dark matter dominated} MFL
systems. Also, in contrast to many
previous efforts, our models are directly compared to the most constraining,
extant data set now available, that for the Carina dSph.
While Carina resembles most other dSphs both structurally and dynamically,
it offers the advantage that it now has the most extensive spatial and kinematical 
coverage (\citealt{Mu06a}) for any dSph outside the Sgr system.
We specifically do not address the Sgr dSph because it has already been explored with similar
methods and models by \citet{Law2005}, and because Sgr has often been argued to be the
{\it exception}, rather than the archetype of 
Galactic dSphs (e.g., \citealt{Mateo1998}; \citealt{Gilmore2004, Gilmore2007}).  

In \citet{Mu06a} we presented empirical findings and arguments that Carina must be 
forming tidal tails, based on the appearance of the satellite at large radii (e.g., Fig.\ 1), 
the implied mass of the system if the outermost Carina stars are bound, and the 
observed velocity shear along Carina's major axis (Fig.\ 2d).
The results of our modeling here lend considerable support to the hypothesis
that the Carina system is presently tidally disrupting.
We show that initially equilibrium, Plummer model satellites on very shocked, radial 
orbits are transformed into systems with properties remarkably similar to Carina (and other dSph) 
satellites.  In these models, the observed extended populations and the flat dispersion profiles
are both a product of tidal debris, despite the presence of dark matter (in an MFL
configuration).

While we do not claim to have a {\it unique} model to describe the nature of the Carina system,
our results dispel several common misconceptions about dSph galaxies
and tidal disruption often invoked to dismiss the tidal disruption hypothesis: We find that
(1) despite the current trend of relying on ever more complex structural
models (i.e. more free parameters) to model dSphs, 
{\it simple, one-component, tidally disrupting satellite models can adequately 
reproduce the strong physical constraints provided by existing observations of at least some 
dSph galaxies}.
(2) Our well-fitting tidal disruption model for Carina {\it even matches the central velocity 
dispersion and $M/L \sim 40$ of the dSph}.  
The presence of dark matter and tidal disruption are not mutually 
incompatible --- {\it MFL, dark matter dominated systems can produce
tidal debris streams that are consistent with current observations}.  This has 
already been suggested by our similar modeling 
of Sgr (\citealt{Law2005}) and Leo I (\citealt{Sohn2007}), but now we demonstrate the 
principle again with a high central $M/L$ dSph system.
Thus, attributing ever larger $M/L$ and extended dark matter halos to dSphs to explain 
their observed outer structure and dynamics  may not be necessary.
(3) Despite considerable periodic shocking, significant tidal disruption, and the presence of
unbound stars near the dSph core, it is found the central velocity dispersion of the 
satellite still gives a reasonable estimation of its bound mass using the traditional
core-fitting methodology.  The influence of 
tides is primarily relegated to the outer parts of the dSph, where the tidal force is 
strongest (see also \citealt{Mashchenko2006}).
(4) The presence of even a small fraction of unbound stars can influence the
observed dynamics of dSphs at large radii.
(5) Sgr may not be a unique example of a tidally disrupting dSph among present MW
satellites.  We conclude by explaining how tidally disrupting systems that are well represented
by an MFL paradigm are not necessarily at odds with the current $\Lambda$-CDM paradigm.

\begin{figure}
\plotfiddle{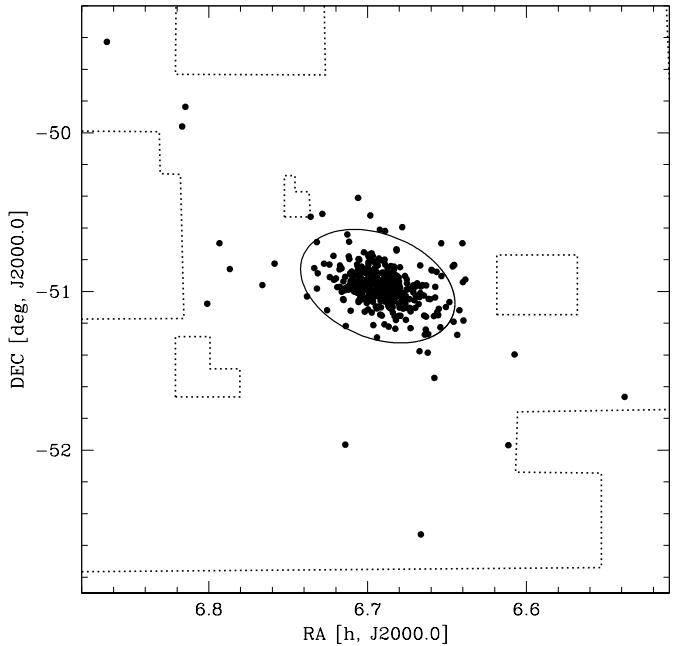}{3.50in}{0}{46.}{46.0}{-140}{-62}
\caption{Distribution of bona fide Carina RV-members on the sky from \citealt{Mu06a}.}
\end{figure}

An ancillary goal of the paper is to demonstrate a methodology to navigate 
the parameter space of satellite orbit, mass and density 
to find matches to the observed structure and dynamics of specific 
dSph galaxies having dynamical and structural data to large radii using 
one-component, MFL, $N$-body simulations.
In this case, we apply the methodology to the Carina system, but it is valid
for application to any other system with sufficient structural and dynamical data.  
A second demonstration of the utilization of this methodology has been given in our study of the 
Leo I system (\citealt{Sohn2007}).

\section{$N$-body Simulations: Model Parameterizations}

Several properties of Carina we have observed 
(\citealt{PaperII,PaperVI}; \citealt{Mu06a}) ---
its substantially extended radial density profile,
the highly elongated
distribution of RV members, the flat velocity dispersion profile --- are consistent
with expectations for a satellite galaxy undergoing tidal disruption.
But this interpretation is perhaps not readily acceptable or satisfying until a
legitimate model or simulation can demonstrate that it is physically possible.
Our goal here is to determine under what circumstances 
(i.e. satellite mass, orbit and density distribution) --- if any  --- the observed Carina
velocity and structural characteristics
can be imitated by simple, MFL 
satellites including tidal effects. 
We have modeled the Carina dwarf
with $10^5$ particles initially in
a \citet{Plummer1911} configuration and with the ensemble orbiting within a static, smooth MW 
potential. The self-gravitating satellite particles have their mutual
interactions calculated using a self-consistent field code (\citealt{Hernquist1992}).  
The form of the adopted MW potential, consisting of a \citet{Miyamoto1975} 
disk, Hernquist spheroid and logarithmic halo, as well as the workings of the simulation 
code is the same as that described in \citet{Law2005}. 
The values for the adopted potential parameters are those used in the  \citet{Law2005}
simulation of satellite disruption that provided the 'best fit' to the spatial 
and velocity distribution observed for tidal debris associated with the Sagittarius 
dwarf spheroidal galaxy. 
In particular, the adopted halo potential is mildly oblate, with minor-to-major axis 
ratio of $q=0.9$. 

To navigate the larger parameter space of satellite mass, density and orbit, we 
first ran a sampling
``grid" of satellite parameterizations to zero in
on the family of models that best matches the observed Carina properties (\S3.1).  We then
experimented with ``fine tuning" models near the ``best-matching" part of parameter space 
in an attempt to improve the match to Carina (\S3.2).

The properties of our initial baseline
grid of models include the following considerations:

(1) Current satellite position and velocity: The model Carina is placed on orbits that yield a 
present Galactic location, heliocentric distance $R$ and systemic RV for the satellite similar 
to that observed for Carina ($l=262^{\circ}$.1, $b=-22^{\circ}$.2, $R=101$ kpc 
and $v_{GSR}=8.0$ km s$^{-1}$).

(2) Satellite mass: In our first tests 
Carina is represented with initial masses of either $10^6$, $10^7$, 
or $10^8$ M$_{\sun}$. 

\begin{deluxetable*}{ c l r r c c r c c}
\tabletypesize{\scriptsize}  
\tablewidth{0pt}
\tablecaption{Properties of Model Orbits }
\tablehead{\colhead{orbit} &
 \colhead{$v_{par}$} &
\colhead{$R_{peri}$ (kpc)} &
\colhead{$R_{apo}$ (kpc)}  &
\colhead{$T_{rad}$ (Gyr)} &
\colhead{$v_{peri}$ (km s$^{-1}$)} &
\colhead{$T_{enc}$ (Myr)} &
\colhead{$M_{peri}$ (M$_{\sun}$ )} &
\colhead{$\rho_{peri}$ (M$_{\sun}$ kpc$^3$)} }
\startdata
1 & 0.2  & $<10$ & $>145$ &  (not stable)&       &            &                                 & \\
2 & 0.3  & 15   & 103  &             1.46 & 360-440  &  32.6  &  1.90$\times10^{11}$   & 2.47$\times10^6$ \\
3 & 0.4  & 22   & 103  &            1.56  & 310-380  &  55.3  &  2.41$\times10^{11}$   & 1.22$\times10^6$ \\
4 & 0.5  & 34   & 103  &             1.67 & 250-350  &  92.8  &  3.29$\times10^{11}$  & 4.96$\times10^5$ \\
5 & 0.6  & 48   & 103  &             1.85 & 245-285  & 160.9 &  4.50$\times10^{11}$   & 2.41$\times10^5$ \\
6 & 0.7  & 64   & 103  &             2.05 & 222-250  & 244.6 &  5.39$\times10^{11}$   & 1.32$\times10^5$ \\
7 & 0.8  & 85   & 103  &             2.28 & 210-227  & 357.8 &  6.81$\times10^{11}$   & 7.41$\times10^4$ \\
8 & 0.9  & 95   & 124  &             2.71 & 198-222  & 408.9 &  7.48$\times10^{11} $  & 5.91$\times10^4$ \\
9 & 1.0  &103  & 152  &             3.06 & 218-236  & 417.0 &  8.02$\times10^{11}$   & 5.01$\times10^4$
\enddata
\end{deluxetable*}

(3) Orbital plane: Based on previous (e.g., \citeauthor*{OLA95}) and our own preliminary 
models of Carina-like systems that demonstrate satellite elongation by tides,
we set the Carina orbital 
plane by the position angle of the satellite 
(65$^{\circ}\pm5^{\circ}$, or roughly 172$^{\circ}\pm5^{\circ}$
in Galactic coordinates, from \citealt{IH95}).  Whether or not this is true, Figure 1 shows that 
at least empirically the more extended Carina RV members do lie primarily along the
central Carina position angle.
Preliminary 
simulations of satellites orbiting with a variety of orbital poles were run, and two (opposite) 
orbital poles were found to satisfy the generation of debris at the observed PA, namely models
with a pole of
$(l,b)=(159.3,-12.9)^{\circ}$ or its antipode at $(339.3,12.9)^{\circ}$. 
Given the symmetry of our model potential and the likelihood that Carina is at its 
apoGalacticon (see below), we 
expect similar results with model satellites run on orbits
with either of these poles, and elect to run the bulk of our model satellites on orbits with
the first pole.
Once we find a favored model with this pole, we run a model with the same overall
parameters but with the opposite pole (\S3.2) to verify the symmetry of solutions.  At this point we 
do not invoke the measured proper motion of Carina by \citet{Piatek2003}, since the relative errors in
this measurement are large and do not well constrain the direction of motion of Carina
(the 95\% confindence interval allows a nearly 80$^{\circ}$ spread in orbital inclination);
however we point out some general consistencies of our best fitting model with 
Piatek et al. results in \S3.2.3.

(4) Orbital shape:  Because the Carina velocity in the Galactic Standard of Rest
convention ($v_{GSR}$) is nearly zero, Carina must 
presently be near apogalacticon, 
near perigalacticon, or be on a nearly circular orbit.  
Given that the satellite radial velocity must be fixed to that observed, the ellipticity of 
the model orbits is set by the adopted transverse velocity, $v_{trans}$ 
for a satellite at the given distance.  
We have created models characterized by $v_{par}=v_{trans}/v_{circ}$ from 0.3 to 1.0, 
where $v_{circ}$ is the Galactic circular velocity at the 
Solar Circle, taken to be 210 km s$^{-1}$
 (see \citealt{Law2005}). The parameter $v_{circ}$ sets the scale of the Galactic mass profile according
to the adopted potential.
The resulting orbits correspond to (peri/apogalactica) of (15/103) kpc to (85/103) kpc for orbits 
where Carina is presently at apogalacticon (orbits 1-7 with $v_{par}\le0.8$).
Orbit 8 with $v_{par} = 0.9$ is relatively circular and orbit 9 with $v_{par} = 1.0$
has Carina near perigalacticon (in an orbit with a 152 kpc apogalacticon).  
Table 1 summarizes the main properties for these orbits, 
including peri- ($R_{peri}$)
and apogalacticon ($R_{apo}$) distance,
radial period ($T_{rad}$), range of peak velocities observed at 
perigalacticon ($v_{peri}$), the
timescale of the tidal encounter at perigalacticon ($T_{enc} =  R_{peri}/v_{peri}$), 
the mass of the MW interior to perigalacticon radius ($M_{peri}$)
and the Galactic density ($\rho_{peri}$) at that radius in our potential.  
We attempted orbits with $v_{par}$ as low as 0.2 (``orbit 1"), 
but these have very small perigalactica that 
cause the orbit to be scattered by the disk potential so that 
no "typical" orbital properties can be given.

(5) Simulation length and time sampling:  For the initial set of models, 
Carina's orbit is traced backwards with a single point mass for the lesser of either five radial 
orbits or 10 Gyr,
and then is evolved forward from this point in full $N$-body mode.  We have run most models
with 4000 even time steps, but have found that the high perigalacticon velocities for some of 
the more radial orbits led to conservation of energy problems at perigalacticon, and finer time
sampling was implemented as necessary.  
Energy was conserved at the $\sim1\%$ level (of the internal satellite energy) 
when orbit 2 models were run with 64000
time steps, when orbit 3 models were run with 16000 time steps and when all rounder orbits
were run with 4000 time steps.

(6) Satellite density:  Apart from the mass, we vary the physical scale 
length, $r_{\rm 0}$, of the model satellite distribution, which is initially 
set to generate a \citet{Plummer1911} model:
\begin{equation}
        \Phi=-{GM_{\rm Carina,0} \over \sqrt{r^2+r_{\rm 0}^2}}
\label{PlummerEqn}
\end{equation}
The density of the satellite, critical to the degree of robustness of the satellite 
against Galactic tidal
forces and therefore a primary driver in the satellite disruption rate, 
is set by the initial mass, $M_{\rm Carina,0}$, and scale, $r_{0}$ of the satellite.
To maintain a constant mass density, the physical scale is sized nominally as
\begin{equation}
        r_{\rm 0}=0.9 (M_{\rm Carina,0}/10^9 M_{\sun})^{1/3}.
\label{ScaleEqn}
\end{equation}

\noindent However, to vary the disruption rate, we adjust the scale by up to a factor of five
(i.e., density varied by more than a factor of 100) at each adopted 
satellite mass.  In all cases, the initial Plummer-configured satellite is allowed to
equilibrate outside the MW potential before being introduced to the
potential at the apogalacticon of the orbit. 

In summary, our initial grid of 80 models has been constructed with three different satellite masses, 
with three or four 
different scale lengths per mass, and with eight different orbital shapes (orbits 2-9).  
Further explorations of satellite parameter space (\S3.2) include additional combinations
of mass, scale, number of radial periods run in the potential, and number of time steps 
in the integration. These extra models pertain primarily to orbit 2 and 3 models (which are found to give the
best matches to the data, as shown below).  In the end, nearly 200 different
$N$-body simulations to model the Carina dwarf galaxy were run on a nearly
devoted Sun Blade 1500 workstation, requiring approximately 6, 24, 
or 96 hours to run simulations having 4000, 16000 and 64000 time steps, respectively.
Table 2 summarizes most of the combinations we have run by
scale and mass, with each table entry indicating which orbit shapes were 
run for a particular mass and scale (and with the bold-faced entries representing the initial
parametrization grid of 80 models).  
Asterisks indicate when a combination of mass, scale and orbit
were run for different numbers (e.g., 1, 2, 3, 4, 5) of total radial orbits. 

\begin{figure}
\plotfiddle{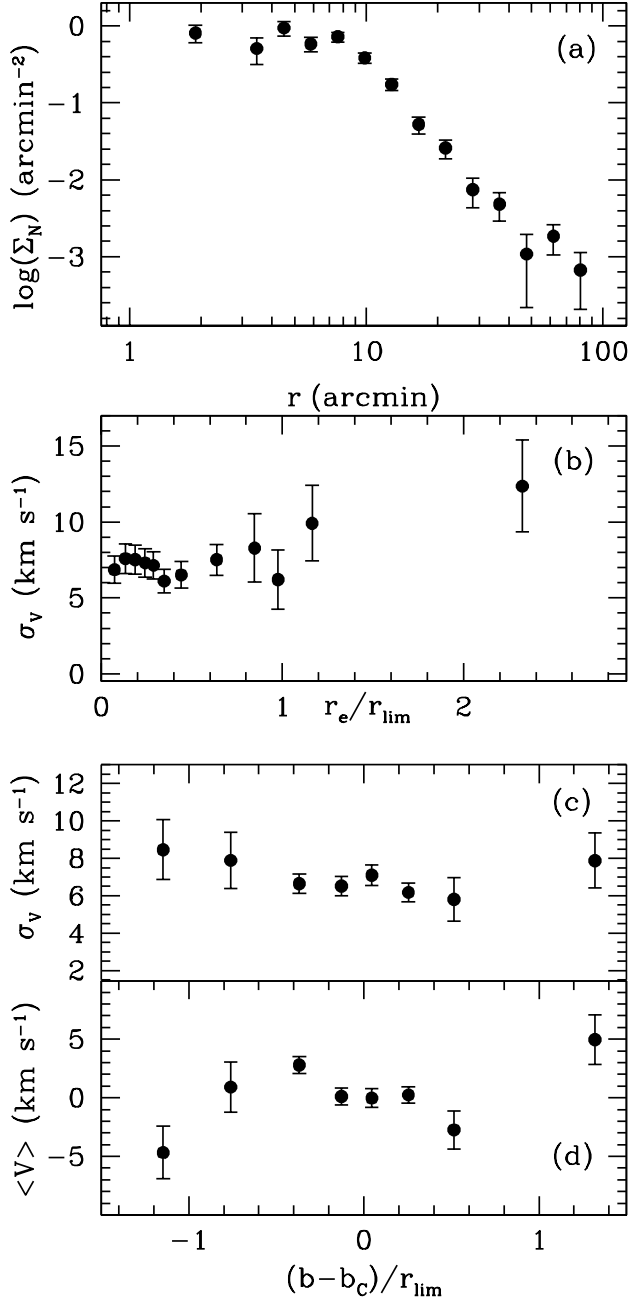}{7.00in}{0}{95.}{95.0}{-285}{-150}
\caption{
 (a) Number density profile for the Carina dSph.
 (b) Observed velocity dispersion as a function of radial distance from the
 Carina's center, normalized to $r_{lim}$.
 (c) Observed velocity dispersion as a function of Galactic latitude, $b$, which
 is a good proxy for the major axis of Carina.  We shift the abscissa to be centered on the
 Galactic latitude of the center of Carina, $b_c$.
 (d) Observed mean projected velocity as a function of major axis distance. All data
 from \citet{Mu06a}.
}
\end{figure}

\section{Results}
\subsection{Results from the Initial Grid of Satellite Parameterizations}

For analysis of our initial grid of simulation parameterizations 
we adopt as primary discriminants for the ``success" of a model its 
ability to match simultaneously
(1) the observed velocity dispersion trend as a function of major axis and radial 
distance from satellite center, (2) the 
general shape and size of the Carina radial density profile, and (3) the mean projected 
velocity along the major axis of Carina. Figure 2 shows these observed constraints,
with data taken from \citet{Mu06a}.  

However, before moving directly to the best fitting models for Carina (see \S3.1.4), we take 
advantage of the fact that our systematic exploration of satellite parameter space 
leads to results that are useful in a more general context,
revealing trends in the nature of satellite tidal disruption that are helpful 
for understanding the observational properties of tidal debris and disrupting satellites. 
To keep the scope of this more general discussion to a managable level, we limit our
discussion primarily to the nature of the models as they appear at the 
distance of Carina, which 
is at apogalacticon for orbits 1-8 and at perigalacticon for orbit 9.  We make this simplification
because Carina's distance
is more or less typical for the classical Galactic dSphs, and satellites on elliptical orbits
spend more time near apogalacticon than in any other orbital phase.

\begin{figure*}
\plotfiddle{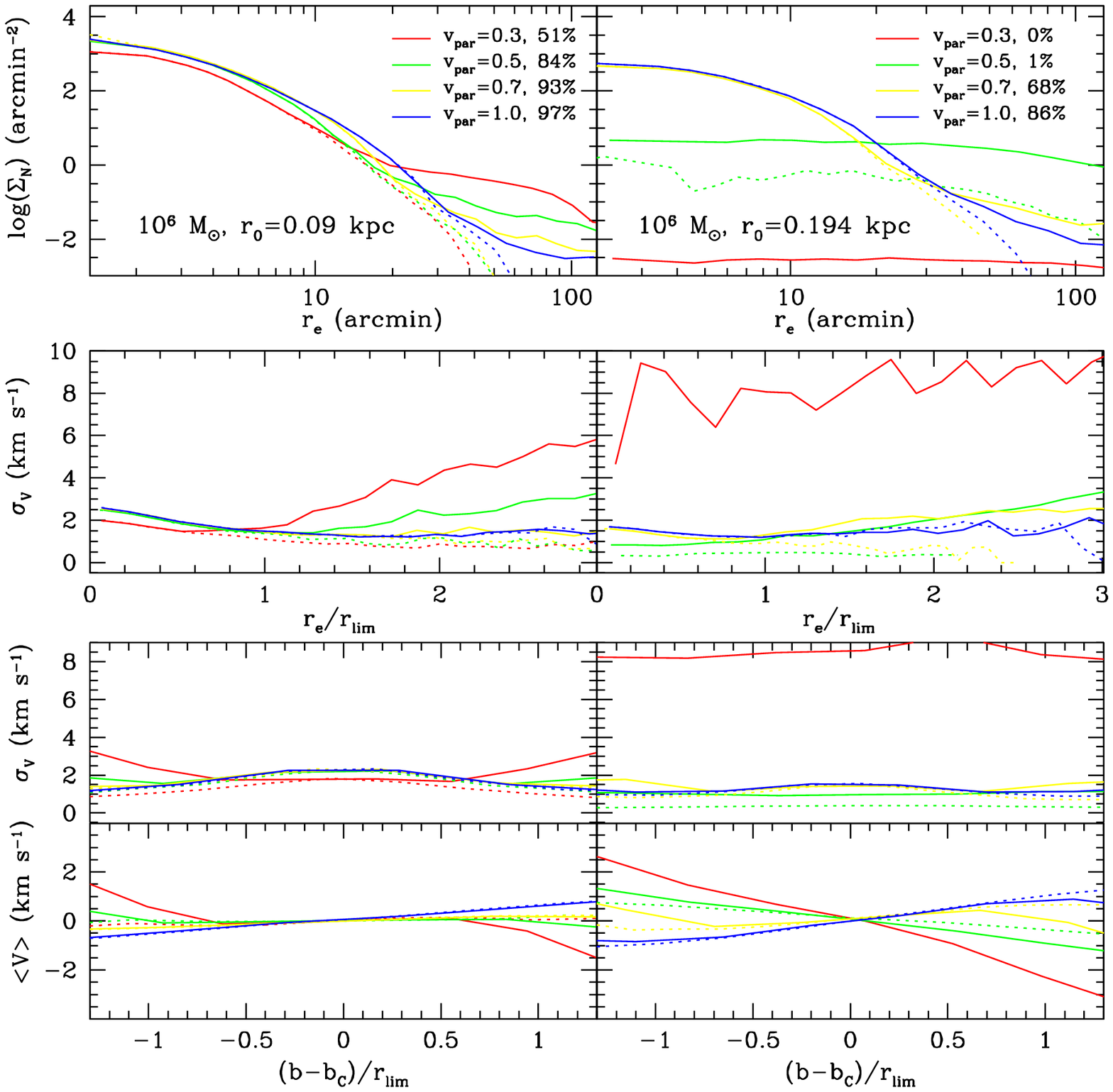}{5.50in}{0}{75.}{75.0}{-235}{-120}
\caption{From top to bottom: Projected number density profiles, velocity dispersion profiles
as a function of radial distance and distance along the major
axis (characterized as a function of Galactic latitude and normalized to the respective $r_{lim}$) and
mean projected velocity along the major axis for ({\it left panels}) model
satellite of mass $1\times10^{6}$ M$_{\sun}$ and scale length of r$_{\rm 0}$=0.090 kpc, in different orbits
($v_{par}=v_{trans}/v_{circ}$=0.3, red lines; 0.5, green lines; 0.7, yellow lines and 1.0, blue lines), and
({\it right panels}) model satellite of mass $1\times10^{6}$ M$_{\sun}$ in the same
orbits as the left panels, but this time with scale length of r$_{\rm 0}$=0.194 kpc.
The percentage numbers represent the fraction of bound mass remaining at the end of the simulation.
Solid lines represent all particles while dotted lines show the behavior of the bound
particles only.}
\end{figure*}

\subsubsection{General trends}

To illustrate the general properties of the model satellites in our initial grid we
have elected to show, first, the salient features of three satellites of the same 
mass, $1\times10^{6}$ M$_{\sun}$, but of different scale lengths (or, equivalently, 
densities) evolved in our Galactic potential. As we have done for the real
Carina data in Figure 2, we show in Figure 3 for the models with two different densities
(a) the projected number density profile, the 
velocity dispersion profiles as a function of both (b) radial distance from the center
of the satellite and (c) position along the major axis (normalized to $r_{lim}$),
which we characterize as a function of Galactic latitude, $b$ (since this is a good
approximation to the major axis for Carina's position angle), and (d) the mean projected
velocity along the major axis.
These properties are shown for four different orbital shapes, from very
eccentric ($v_{par}=0.3$) to rather circular ($v_{par}=1.0$).  In all models
at least some degree of tidal disruption occurs and tidal debris dominates 
the structure and dynamics at large radius.

Some primary sweeping trends are demonstrated in Figure 3. For example, the central velocity
dispersion remains almost unaltered by tides (in the sense 
that its value does not inflate) until the satellite is near complete destruction whereas, 
at large radii, the velocity dispersion highly 
correlates to the degree of tidal stripping (see also Fig.\ 6). 
In fact, as found by \citeauthor*{OLA95}, as long as the satellite retains a predominantly bound core, the 
value of the central velocity dispersion correlates with the 
instantaneous bound mass of the satellites. In the cases where the system is
completely disrupted (i.e., indicated in the figure by having 0\% bound mass today), a totally flat velocity 
dispersion profile is observed, and the magnitude of the
dispersion is no longer correlated to the starting mass, but is also
influenced by the initial satellite density and orbital shape. 
Density distributions, on the other hand, seem to be, for the most part, only mildly affected by 
the degree of tidal disruption
when compared to the velocity dispersion profiles for a given mass/density/orbit combination. 
The main effect of tidal disruption on the density profiles is for the dSphs
to develop an extended component that ``breaks" from the initial
Plummer configuration, as has already been found in previous studies (e.g., \citealt{JZSH99};
\citealt{Read2006b}; \citealt{Klimentowski2007}). An overall enlargement of the satellites occurs only when
the systems are near destruction, after which they lack a core in the density profile and 
present a rather flat spatial distribution.

\begin{figure}
\plotfiddle{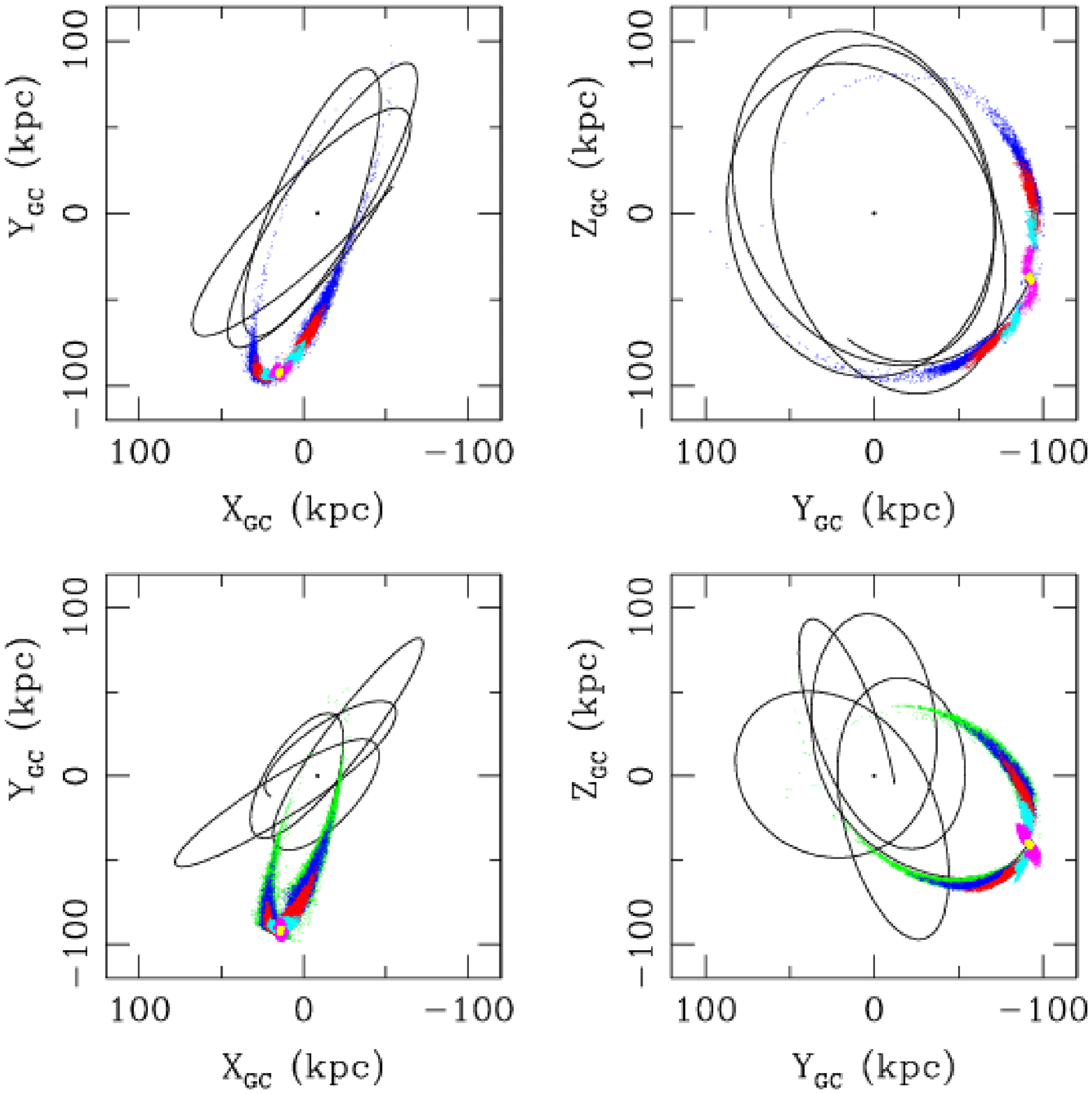}{3.70in}{0}{46.}{46.0}{-130}{12}
\caption{$X_{GC}-Y_{GC}$ and $Y_{GC}-Z_{GC}$ Cartesian projections for a model
with a starting mass of $10^7 M_{\sun}$ in two rather benign orbits. The upper panels
show the satellite in a $v_{par}=0.8$ orbit while the lower panels show the system in a
$v_{par}=0.5$ orbit. The coloring scheme is as follows:
Yellow dots mark particles that are still bound, while green, blue, red, cyan and magenta
represent material that became unbound after 1,2,3,4 and 5 orbits respectively.}
\end{figure}

The general appearance of the debris for low eccentricity orbits is illustrated in Figure 4,
where we show the $X_{GC}-Y_{GC}$ and $Z_{GC}-Y_{GC}$ Cartesian projections (centered on the 
Galactic center) for a model with an initial mass of $1\times10^{7}$ M$_{\sun}$
in two different, rather benign orbits. 
Tidal debris released at different times is represented by different colors, with the yellow
particles marking the presently bound population.
The upper and lowel panels show the model in a
$v_{par}=0.8$ (nearly circular) and $v_{par}=0.5$ (3:1) orbit respectively.  In both cases, rather
narrow, coherent debris streams are formed.

In constrast, Figure 5 shows the $X_{GC}-Y_{GC}$ and $Y_{GC}-Z_{GC}$ Cartesian projection of two
$1.0\times10^{7}$ M$_{\sun}$ satellites of different densities in a very radial
orbit ($v_{par}=0.3$ and $R_{peri}=15$ kpc) after five perigalacticon passages.
The upper panels show a model with a scale of $r_{\rm 0}=0.280$ kpc (model 142) and the
lower ones a model with $r_{\rm 0}=0.194$ kpc (model 102) --- i.e. lower and higher density respectively.
In both cases the debris is more broadly dispersed, but while
model 102 retains a predominantly bound core after five orbits, model 142
has been completely disrupted and lacks a core in the density profile 
(a property observed in all of our model satellites that have been completely destroyed).  
As Figures 4 and 5 demonstrate, for a given orbit and mass, the mass-loss-rate is driven by the 
density of the satellite (\citealt{JHB1996}; see also Figure 6, where the density is kept constant 
and the mass-loss-rate stays the same for a given orbit, even as mass is varied).

\begin{figure}
\plotfiddle{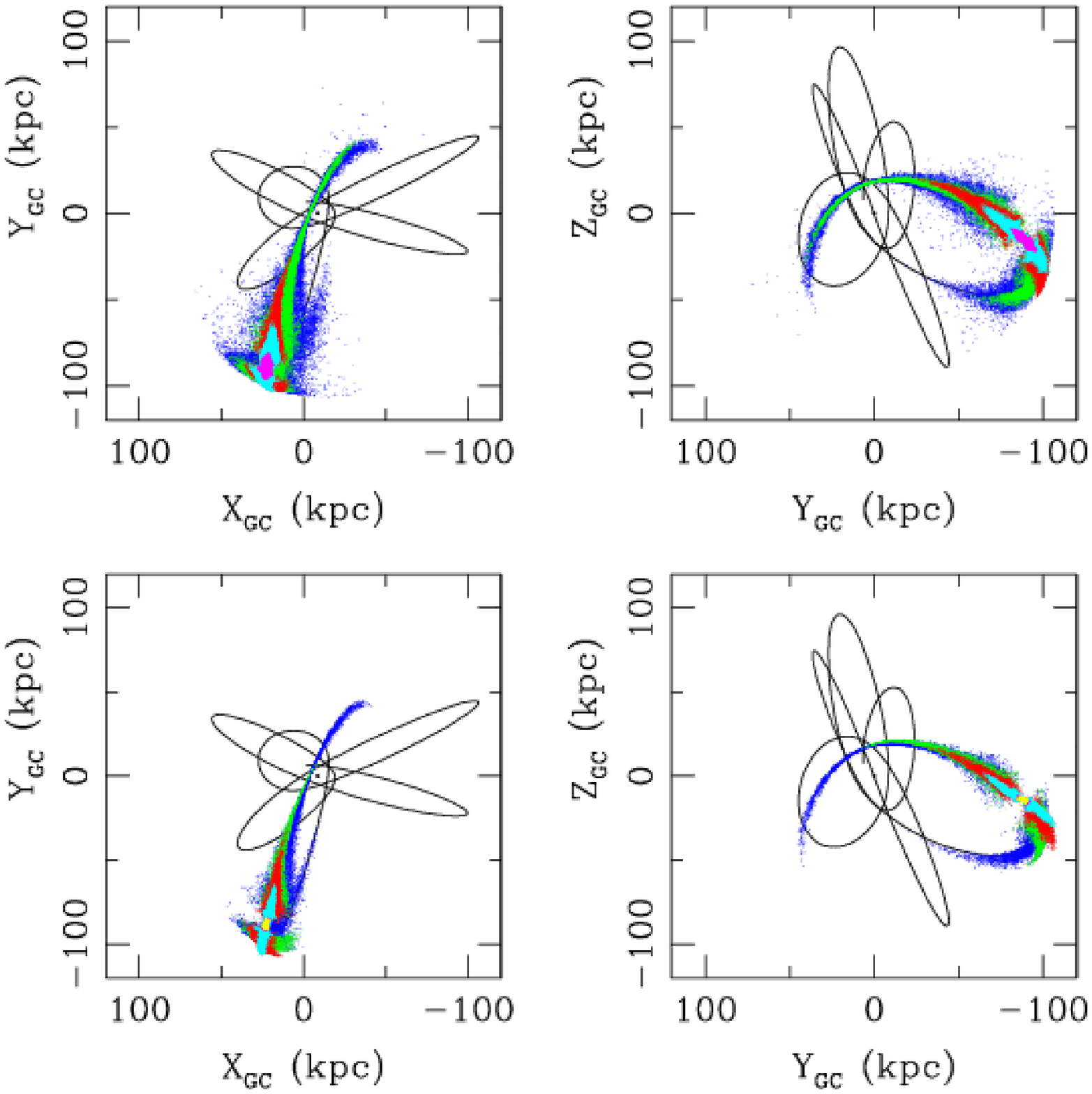}{3.70in}{0}{46.}{46.0}{-130}{12}
\caption{$X_{GC}-Y_{GC}$ and $Y_{GC}-Z_{GC}$ Cartesian projections for two models in rather
eccentric orbits ($v_{par}=0.3$) after five perigalacticon passages.
Both models had an initial mass of $1\times10^{7}$ M$_{\sun}$ but
the upper panels show the model with a $r_{\rm 0}=0.28$ kpc (less dense)
while the lower panels show the model with a $r_{\rm 0}=0.194$ kpc (more dense).
The coloring scheme as the same as in Figure 5.
As may be seen, the lower density model is destroyed after five orbits (no
yellow component) but the higher density model keeps a predominantly
bound core.}
\end{figure}

Other basic trends depend heavily on the orbit shape, e.g., 
more circular, rather gentle orbits (apo/perigalacticon ratio no greater 
than 3:1) versus more energetic, rather eccentric orbits. These results are 
summarized in Figure 6, which shows the density, mean velocity
and velocity dispersion trends for three disrupting satellites of different masses
($1\times10^{6}$, $1\times10^{7}$ and $1\times10^{8}$ M$_{\sun}$)
in two different orbits ($v_{par}=0.8$, left panels; $v_{par}=0.3$, right panels), this time 
with central density kept constant (by adjusting the satellite scale to track the cube root of the
mass according to Eq.\ 2) in order to explore the trends with mass.

After five orbits (or, alternatively, 10 Gyrs, see \S2.1), 
most of the initial mass still remains bound for model satellites in gentle orbits 
(left panels of Figure 6), 
regardless of their initial masses.
At the same time, all of the simulated satellites in these orbits experienced some degree of tidal stripping 
that leads to extended spatial components populated by tidal debris.
The left panels in Figure 6 
show that model satellites with large perigalacticon distances and almost circular
orbits ($R_{peri}$=85 kpc and $v_{par}=0.8$) 
undergo relatively modest tidal stripping and
their velocity dispersion trends are found to {\it decline} over degree (kpc)
scales, as expected for mostly bound systems (containing particles on predominantly non-circular
internal orbits) where the velocity dispersion is the highest at the center and then decreases 
at large projected radii.
However, the velocity dispersion at large radii still departs from the initial, unperturbed value,
indicating some tidal heating of the bound population at large radii.

\begin{figure*}
\plotfiddle{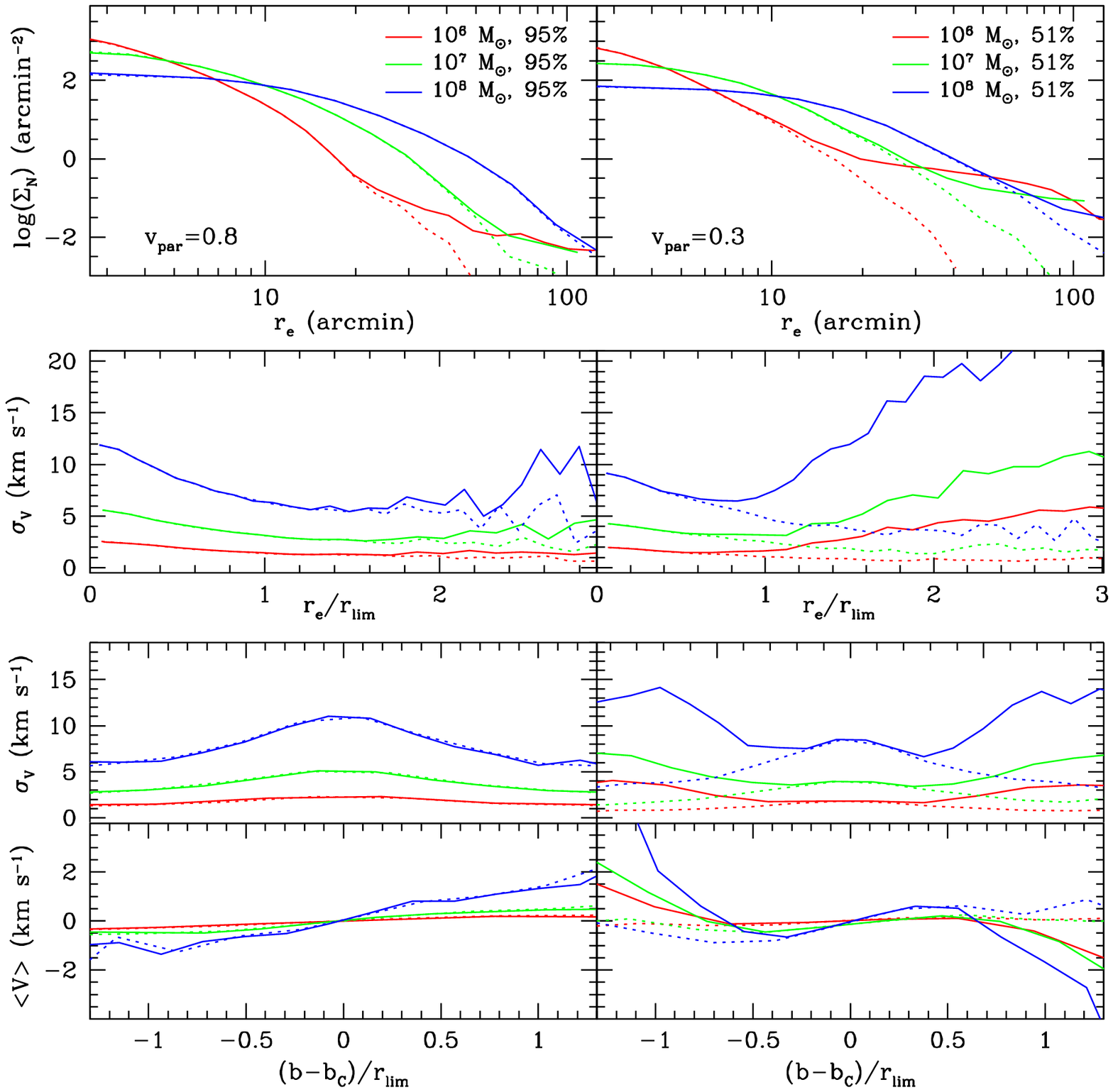}{5.50in}{0}{75.}{75.0}{-235}{-120}
\caption{Same as Figure 3 but this time,
({\it left panels}) model satellites of masses $1\times10^{6}$ (red lines), $1\times10^{7}$ (green lines)
and $1\times10^{8}$ M$_{\sun}$ (blue lines)
with initial central density kept constant by adjusting the Plummer scale length (see text) in a rather
circular ($v_{par}=0.8$) orbit, and ({\it right panels}) the same model satellites but now in a eccentric
orbit ($v_{par}=0.3$). As in Figure 3, the percentage numbers represent the fraction of bound mass
remaining at the end of the simulation.
Solid lines represent all particles while dotted lines show the behavior of the bound
particles only.}
\end{figure*}

As the orbits become less circular (and the perigalactica smaller), tidal stripping
becomes more vigorous and more unbound particles are produced.
The right panels in Figure 6 show how tidal debris (unbound particles) generally
produce flat or rising dispersion profiles at large radii from the dSph center.
In fact, the size of the RV dispersion in the tails correlates with both
the mass of the satellite and the orbital shape as long as the satellite retains
a predominantly bound core;
this is as was found in the modeling of the Sgr system (\citealt{Law2005}), and was, in fact,
one way that was used to constrain the mass of the Sgr system in that study.

\begin{figure}
\plotfiddle{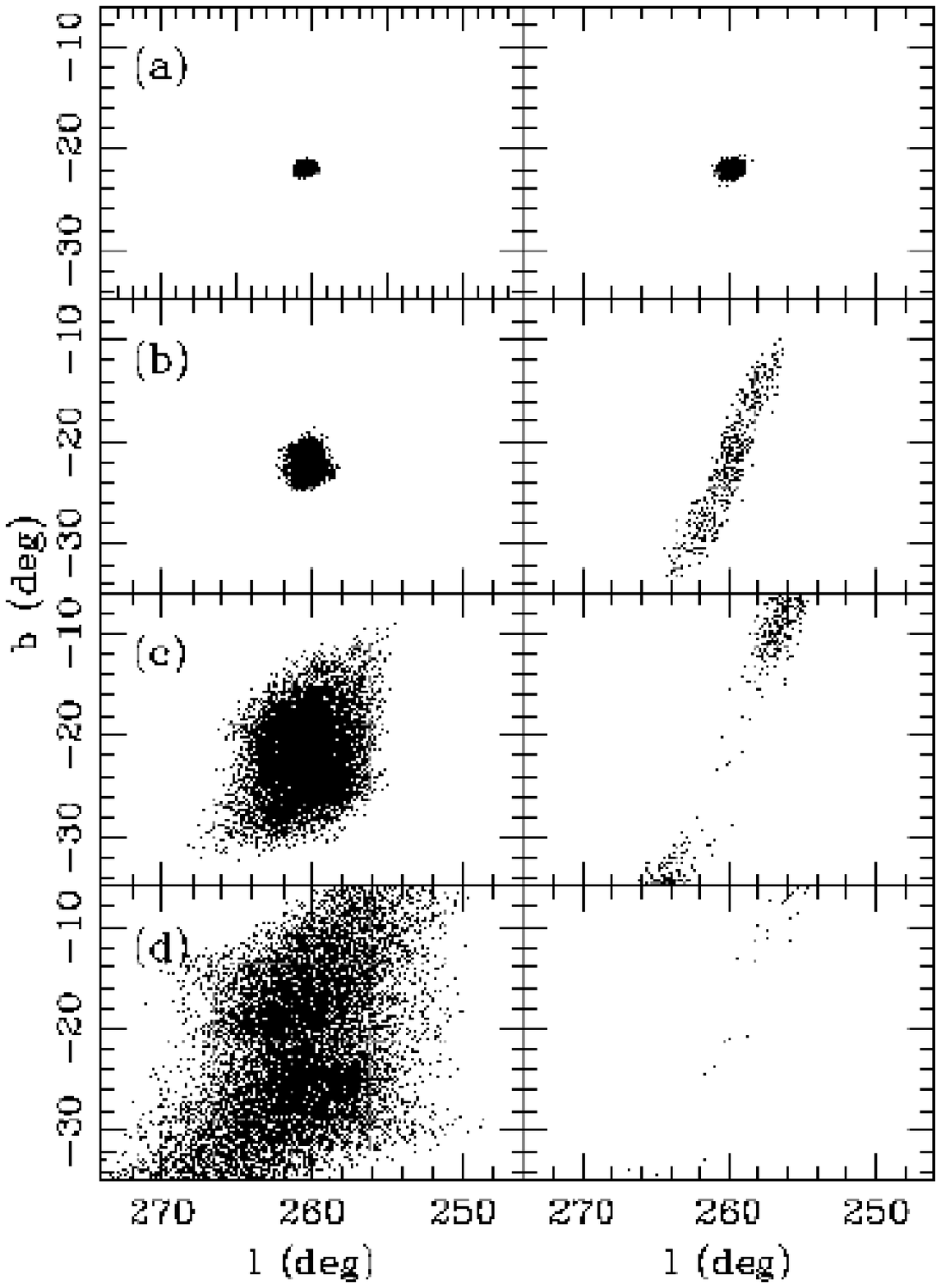}{4.70in}{0}{66.}{66.0}{-200}{-88}
\caption{(a) Spatial distribution of bound particles for a
model of $1\times10^{7}$ M$_{\sun}$ and $r_{0}=0.194$ kpc.
Panels (b), (c) and (d) show the spatial distribution of particles lost during the last
and successively previous orbits respectively (equivalent to the different colors
debris shown in the other models portrayed in Figures 4 and 5). 
Left panels show the satellite in an eccentric ($v_{par}$=0.3) orbit while
the right panels show the same satellite in rather circular ($v_{par}$=0.8) one.
As may be seen, the nearby debris is dominated by particles released
during the last orbit, although debris released in previous passages can
linger near the center for many orbits in the case of eccentric orbits.}
\end{figure}

A key feature demonstrated by these radial orbit models --- indeed by all models
retaining bound cores (e.g., in Figures 3 and 6) 
--- is that the central velocity dispersion is still dominated by the bound population 
even when the outer parts have greatly inflated velocity dispersions. Thus 
Galactic tides have little effect on the inferred central $M/L$,
even in systems with high disruption rates.
However, a radial rise to a {\it higher} velocity dispersion is also a characteristic product
of the unbound component in these radial orbit models.
This rise in velocity dispersion can happen well within
the limiting radius of the bound system, reflecting the presence of unbound
particles quite near the center of the satellite.
To illustrate this, in the left panels of Figure 7 we show the spatial distribution 
(in Galactic coordinates) of particles near the limiting radius for 
a model in an eccentric orbit ($1\times10^{7}$ M$_{\sun}$, $r_{\rm 0}=0.194$ kpc, $v_{par}=0.3$). 
Panels from top to bottom show particles that have become unbound in successive orbits, with
the top panel showing the presently bound population.
This figure demostrates that, in general, the centralized tidal 
debris responsible for inflating the velocity dispersion radially, is 
dominated by the ``puff" of debris released on the last perigalactic
impulse, although particles unbound from prior perigalactica can linger near the dSph
center for many orbits.
For comparison, the right panels of Figure 7 show the same feature but for the model 
satellite in a more benign orbit ($v_{par}=0.8$) where the tidal debris 
dissipates much more quickly.

\subsubsection{Tidally Induced Rotation?}

\citeauthor*{PP95} pointed out that the most unique and readily observable 
consequence of tidal disruption is tidal shear that resembles solid-body rotation 
along the major axis of the dSph. 
To compare our results to these expectations, the lower panels of Figures 3 and 6
show the mean projected velocity along the major axis distance for the same models shown in
the upper panels.
Satellites in nearly circular orbits show, in general, a smooth velocity gradient 
within their $r_{lim}$ (e.g., Fig. 6, left panels). However, this 
small velocity gradient is not a result of Galactic tides (there is only a small degree
of tidal disruption suffered by satellites in these
orbits) but is predominantly a result of the line-of-sight projection of the 
satellite mean velocity along the orbital path.
On the other hand, satellites in more radial orbits do more strongly show
the effect described by \citeauthor*{PP95} (e.g., Fig. 6, right panels). 
While satellites that have been completely destroyed show, for the most part, a smooth gradient along the major axis,
satellites on more radial orbits that retain at least some bound core show a ``rotation" ($S$-shape) 
signature
like that described by \citeauthor*{PP95}.  However, it is important to note that this signature 
is observable mostly outside the $r_{lim}$ of the satellite core, whereas within $r_{lim}$ the 
magnitude of the signal is typically relatively modest --- e.g., small compared to the central 
velocity dispersion.  
In these cases of satellites with bound cores, the bound component shows almost 
no signature of tidal shear and the 
``rotation" is reflected only in the unbound particles (thus, this ``shearing" or "rotation" is really 
just a reflection of 
how the tidal debris with its net angular momentum and energy is organizing in reponse to the MW potential).
These results suggest that, while apparent rotation can be a consequence
of tidal stripping, as suggested by \citeauthor*{PP95}, for satellites that are not completely 
destroyed this effect can occur well outside the typical radii to which most current RV surveys 
of dSphs have been limited.  
Thus, in most cases the \citeauthor*{PP95} signal cannot be counted on as evidence for lack of 
tidal effects since {\it the rotation would generally not be expected to be seen}, or, if detected, 
it would likely be deemed as ``statistically insignificant" in most studies.  
We return to this point in \S4.1.

\begin{figure} 
\plotfiddle{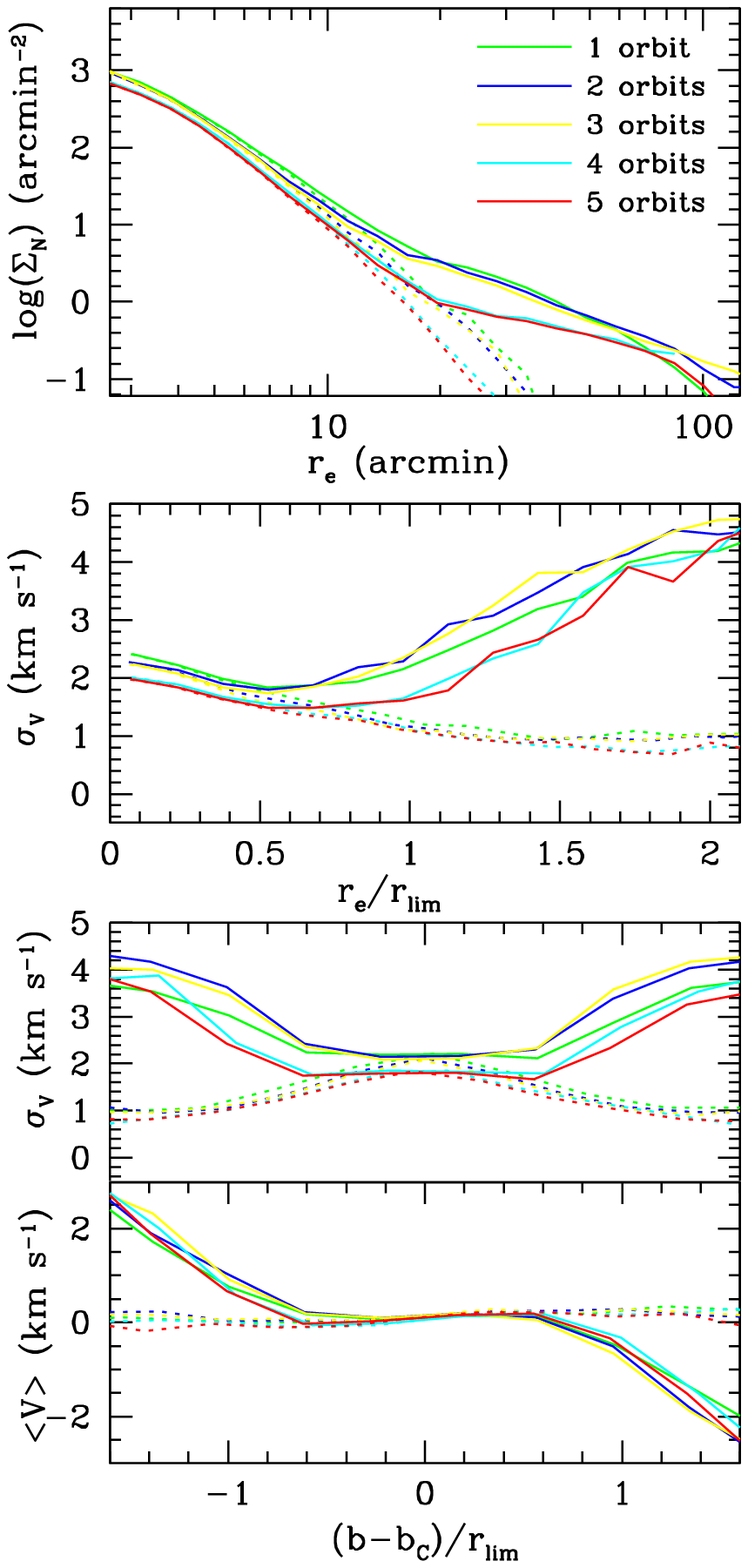}{7.00in}{0}{95.}{95.0}{-285}{-150}
\caption{The evolution of the density, velocity dispersion and mean velocity profiles over
five orbits for one of the low mass, medium density, radial orbit  models showing the
general properties observed in Carina.
This shows that satellites can retain the general character of a rising
velocity dispersion over many orbits, but the asymptotic dispersion
value is a function of the amount of past disruption.}
\end{figure}

\subsubsection{Time Dependence of the Satellite Properties}

As mentioned in \S1.1, a primary problem with previous modeling of tidally
disrupting, low $M/L$ satellites is that any successful matches to observed dSph
satellites depends on catching these systems in one particular phase of evolution.
However, in all cases we have explored, save those where the satellite is completely destroyed, the
radial density profile of the satellite resembles an ``inner King+extended population"
profile that looks nearly the same for different orbital passages when viewed at the
same orbital phase (uppermost panels of Figures 2, 3, 6, 8 and 9).
Figure 8 demonstrates our general finding that the overall spatial and dynamical
character of disrupting satellites can be maintained for long periods of time --- as long as
the satellite maintains a bound core but undergoes a more or less steady mass-loss-rate.
This is an advantageous property of our models over the previous, more specialized
$N$-body simulations of tidal disruption.
In contrast to, for example, the findings of \citeauthor*{PP95},
the specific physical satellite properties we seek in the models are not
rare phenomena that we have to ``catch" at special times in the life of the satellite,
but rather long-lived phenomena that require no invocation of particular
timing (apart from observing the system at the proper orbital phase --- near apogalacticon
in the case of Carina, which is a position near where satellites spend a considerable fraction
of a radial period).
Our dSph models also differ from those of \citet{Kuhn1993}
and \citet{Kroupa1997} where the satellites must be observed at very specific times
in their tidal evolution (namely, after total disruption or just prior to total
disruption for these two references, respectively) and along specific lines of sight.

On the other hand, Figure 8 shows that while the overall character of the 
satellite density and velocity dispersion profiles is maintained for long periods 
of times, there is {\it some} gradual evolution in the scaling of the density profile 
and the magnitude of the velocity dispersion at large radii that is dependent on 
the amount of past disruption.  
Figure 8 shows the velocity dispersion profile observed today for satellites with 
the same initial mass (in this case $10^6$ M$_{\sun}$) on the same radial orbit, but 
which have been dropped into the Galactic potential at successively earlier apogalactica.  
In this case, a central dip to a progressively lower central velocity dispersion is 
observed, while the dispersion of the debris-dominated part of the system changes with time, 
with slighly colder debris accumulating after the satellite has suffered more mass loss.  Of 
course, this mass loss causes the overall density of the satellite (at all radii) to drop 
with time.  The trend of the phenomena shown in Figure 8 thus admit yet another ``free 
parameter" --- i.e. number of orbits of tidal interaction --- that one may tune, in addition 
to orbit shape, initial satellite mass, and satellite scale, to match the properties of 
any particular real satellite, such as the Carina system.

\subsubsection{What is needed to match Carina?}

In only one part of the systemic survey of orbit, mass, density parameter
space do the simulations
reproduce the general character of the observed Carina
velocity dispersion profile {\it as well as} the general projected
density distribution of Carina. 
The operative conditions in common among the ``successful" models is a satellite 
that experiences intense, periodic tidal shocks at the perigalacticon 
of a very radial orbit, but one that
also retains a bound core.  Most of the models run with a $v_{par}=0.3$ satellite
orbit (orbit 2), which have peri:apogalactica of 15:103 kpc and radial periods of
1.5 Gyr, share this character (e.g., Fig. 8), as do some model satellites on 
the less extreme orbit 3 ($v_{par}=0.4$).

We may summarize the results of this initial survey of parameter space to find a 
model that resembles Carina as follows:

\begin{itemize}

\item It is possible to find MFL $N$-body models of disrupting satellites,
constrained to the present position and radial velocity of Carina,
that reproduce the general morphological character (i.e., a ``inner King+extended population" radial 
density profile) and dynamical character (i.e., a flat/rising dispersion profile
with radius) observed in the Carina system.

\item The common properties of models succeeding in this way are that the 
satellite
retains a bound central core
but experiences periodic, impulsive shocks on a relatively radial (e.g., 7:1) orbit that 
spawns nearby clouds of unbound tidal debris.
Carina resembles the system as observed near the apogalacticon of one of these orbits.

\item The rise in the velocity dispersion away from the core is caused by the growing
contribution of unbound debris 
which becomes important even well inside the projected limiting radius of the bound 
component.  The rising dispersion
profile at larger radii is due to the dominance of tidal debris there.

\item 
Our simulations do not depend on special timing to match the general Carina properties, other than
observing the system near the correct orbital phase.  Thus, the results of our
modeling do not hinge on particularly special circumstances other than that the 
satellites be on rather radial orbits.

\end{itemize}

\begin{deluxetable*}{r | c c c c c c c}
\tabletypesize{\scriptsize}
\tablewidth{0pt}
\tablecaption{Model Parameterizations\tablenotemark{a}}
\tablehead{ & \multicolumn{7}{c}{Scale} \\   
\cline{2-8} \colhead{initial mass ($M_{\sun}$)} & \colhead{0.045} & \colhead{0.090} & \colhead{0.194} & \colhead{0.280} &
\colhead{0.350} & \colhead{0.418} & \colhead{0.550} }
\startdata
$10^6$                    &       2        &  {\bf (1-6)*,7-9} & {\bf 2-9}     &                    &              & {\bf 2*,3-9} &  \\
$3{\times}10^6$     &    1,2*     &   1,2*                 &                    &                    &              &                   &    \\
$10^7$                     &                & {\bf (1-4)*,5-9} & {\bf 2*,3-9} & {\bf 2*,3-9} &            & {\bf 2-9}     &   \\
$3{\times}10^7$      &                &                          &                     &      2            &              &                    &   \\
$3.75{\times}10^7$ &               &                           &                    &                     &      2     &                    &    \\ 
$4.5{\times}10^7$   &               &                           &                    &                      &              &      2           &   \\
$10^8$                      &               & {\bf 2-9}            & {\bf 2-9}     &                       &             & {\bf 1-9}    & 2 \\
\enddata

\tablenotetext{a}{Entries give orbits run for each satellite and scale mass. Boldface entries
highlight the initial parameterization grid.  Asterisks denote models run with multiple versions
having the satellite undergo different numbers of orbits (perigalactic impulses) to the present
time.}

\end{deluxetable*}

\noindent In the next section we explore this family of models further, both to find a better 
match to the Carina data by fine-tuning the orbital properties, and to understand the implications
of such models for the possible physical state (e.g., mass, $M/L$, mass-loss-rate) of the Carina system.

\begin{figure}
\plotfiddle{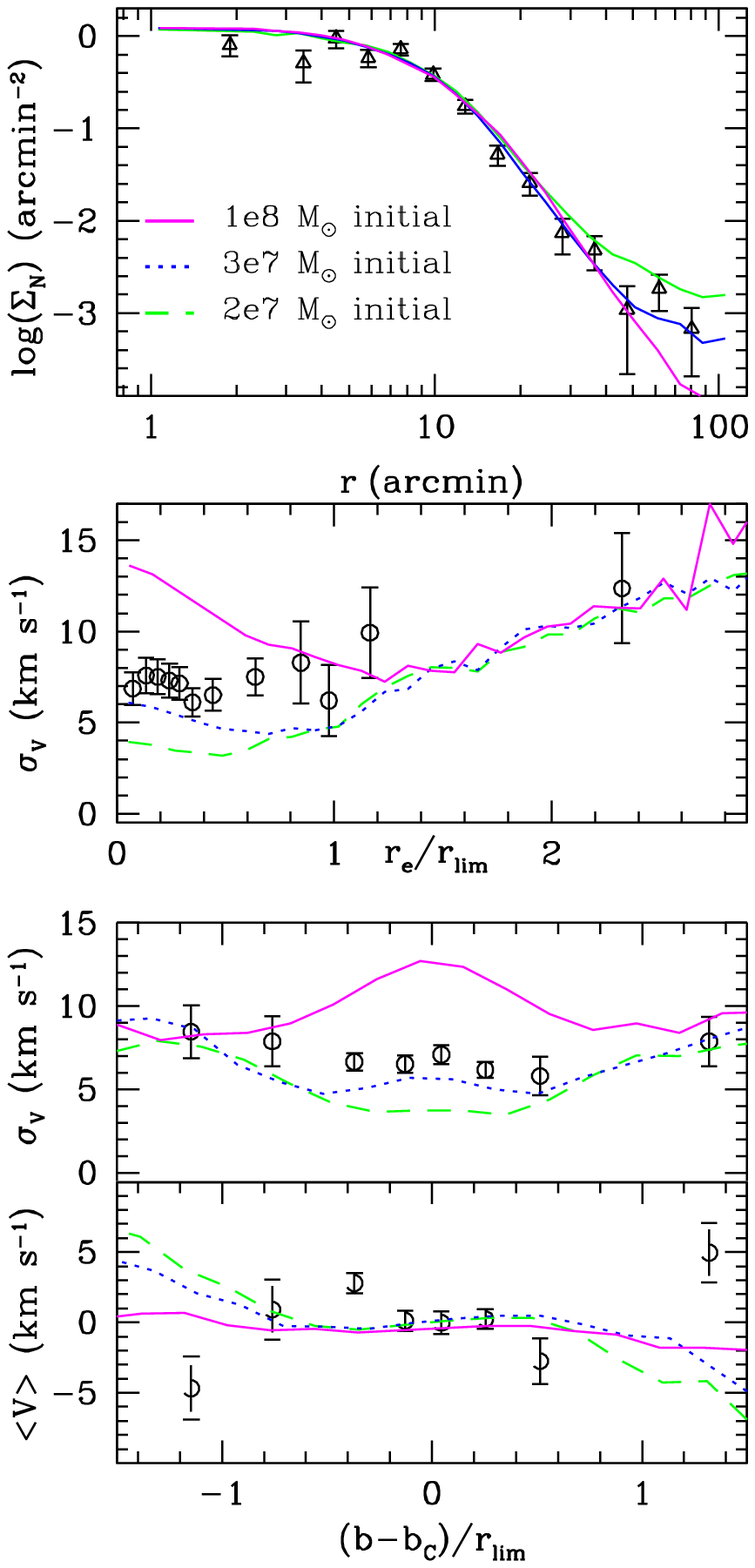}{7.00in}{0}{95.}{95.0}{-285}{-150}
\caption{Three models with an initial physical scale of
$r_{\rm 0}=0.280$ kpc, chosen to yield present satellite sizes
comparable to that of the luminous component of Carina (see text
for details). Three different initial masses of 2,3 and $10\times10^{7}$
M$_{\sun}$ are shown to explore the resulting number density, velocity dispersion
and mean velocity profiles. Data from \citet{Mu06a} are
shown for comparison as open symbols with error bars.}
\end{figure}

\subsection{Fine Tuning the Models}

\subsubsection{Zeroing in on Carina}

>From the previous general survey of single component $N$-body simulations we understand that 
orbital shape plays a critical role
in producing the dynamical effects we seek to emulate in the Carina system, and 
that very radial orbits are preferred for a good match to the observed Carina properties.
To obtain an even better match to the observations, it is worthwhile investigating a 
finer gradation of parameter space for these radial orbit models.  For now, we focus on
models with satellites on orbit 2.  With our single-component models this leaves as free
parameters to vary: the initial satellite total mass, scale, and duration of the numerical 
integration (i.e., number of radial orbits).
These parameters are expected to affect the satellite and debris in certain ways for a 
particular orbit:

\begin{itemize}

\item The measured King core radius ($r_{core}$) and $r_{lim}$ 
will obviously be influenced by the input physical scale of the satellite, $r_{\rm 0}$.

\item The central velocity dispersion of the satellite is expected to vary 
dimensionally as (mass/scale).

\item The amount of debris is set by the mass-loss-rate and this is correlated to the density of
the satellite (for a given orbit).  Thus it scales as (mass/scale$^3$).  This will 
affect not only the relative density of the extended population to the central density 
of the satellite in the radial profile, but, of course,  how quickly the velocity 
dispersion switches from being dominated by bound versus unbound debris.

\item As we have seen above  and in our previous studies
of Sgr debris (\citealt{Law2005}), the velocity dispersion of tidal debris correlates
with the satellite mass as long as the satellite maintains a bound core.

\end{itemize}

\noindent Guided by these rules of thumb, we attempt to vary the mass and scale of the
model satellite in a way to best match the radial velocity dispersion and radial density profiles
simultaneously.   

First, we set the physical scale. Since we are looking for a model that
matches all of the Carina properties, we start by requiring the model satellite
to have a physical scale matching as closely as possible that of Carina.
Considering that the size of the bound component of the model satellites does not 
vary significantly over their lifetime (e.g., Figure 8), 
we set the Plummer physical scale of our model Carina guided by the observed Carina
core radius, $r_{core} =$ 258 pc (\citealt{Mateo1998}). 
Not surprisingly, King distributions fitted to satellites modeled in our initial general survey
(see Appendix A)
show that those with $r_{\rm 0}=0.280$ kpc yield the closest match to the Carina density profile ---
i.e., those for which $r_{\rm 0}$ is similar to Carina's $r_{core}$.\footnote{While $r_{core}$ 
radius and the Plummer length scale of a given density distribution are not strictly 
equivalent, their ratio is of order unity for low concentration systems like dSphs.}

Once the physical scale is set, we seek a model that 
is capable of reproducing the observed {\it central} velocity dispersion 
of Carina ($6.97\pm0.65$ km s$^{-1}$, from \citealt{Mu06a}).
Since this varies as (mass/scale) and the scale of the model
is already set, that leaves us with the initial mass as the pertinent variable.
But by varying the initial mass we are also simultaneously changing the initial
density of the model and that drives the mass loss rate, which, in turn, affects 
the shape of the velocity dispersion profile as well as the current bound mass of the satellite;
these observational constraints {\it each} tightly and independently constrain the choice of mass.
Nevertheless, a model reasonably matching {\it all} of these constraints can be found.\footnote{The
fact that all of these constraints can be simultaneously satisfied lends considerable weight to 
the viability of the single component models; after all, were they an incorrect description 
of the physical state of Carina, it is easy to imagine our models having difficulty satisfying all 
observable constraints with the limited number of free parameters.}
For example, Figure 9 shows the velocity dispersion, projected mean velocity and density 
profiles for satellites in a $v_{par}=0.3$ 
orbit with initial masses of $2\times10^{7}$, $3\times10^{7}$
and $1\times10^{8}$ M$_{\sun}$, after 5 orbits. The lower and higher mass models 
yield current central velocity dispersions that are too low and high compared to Carina respectively,
while the intermediate mass model gives the 
closest match. 
This intermediate mass model provides a better-matching density distribution as well.

While Figure 9 demonstrates that of the three models shown 
the $3\times10^{7}$ M$_{\sun}$ model provides
the closest match to the shape of the velocity dispersion {\it profile}, its values 
are still systematically low compared with the data. Further fine 
tuning of the model Carina thus requires an overall increase of the velocity dispersion.
We achieve this by increasing the initial mass of the model while at
the same time making the orbit slightly more eccentric to keep a similar mass loss
rate. The net effect is a model that yields the right central velocity dispersion while at the same time
providing a good match to the shape of the velocity dispersion and the number density 
profiles (those shown in Figure 10).

Finally, we note that models initially run and presented so far were designed to have an orbital pole
of $(l,b)=(159.3,-12.9)^{\circ}$. 
But when we eventually collected spectroscopic 
data at large radius (presented in \citealt{Mu06a}), we found
that these models were actually yielding
the opposite mean velocity trend from that being found in Carina (Fig. 2). 
To fix this difference with our completed simulations, in the end
we re-computed our best matching models by placing the satellites on orbits with the
polar antipode of
$(l,b)=(339.3,12.9)^{\circ}$ to obtain the correct velocity trend.

Tables 3 and 4 give the parameters for two models (502G and 502I) that yield
close matches to the data.  
As may be seen, these ``best matching" models to the Carina structure and dynamics 
have initial masses of $3.6$ and $3.8\times$10$^{7}$ M$_{\sun}$, but 
have evolved after five orbits to present masses of $\sim1.9\times$10$^{7}$ M$_{\sun}$.  
Interestingly, these final masses are comparable to the 
current mass of Carina derived using the core-fitting method (\citealt{Mateo1993};
\citealt{Mu06a}; see \S3.2.2).
Figure 10 compares the final state of these simulations against the data for Carina.
In the top panels, we compare the number density profiles. In the middle pairs of panels we 
compare the velocity dispersion profiles both as a function of angular distance from 
the center of Carina and as a function of Galactic latitude respectively, and in the bottom 
panels we compare the mean velocity trend of the satellite in Galactic Standard of Rest
velocities. 
For both simulations 502G and 502I, the models lie within 1-$\sigma$ of the 
data for almost all positions in the dSph for {\it all four} of the observational trends shown.

\begin{figure*}
\plotfiddle{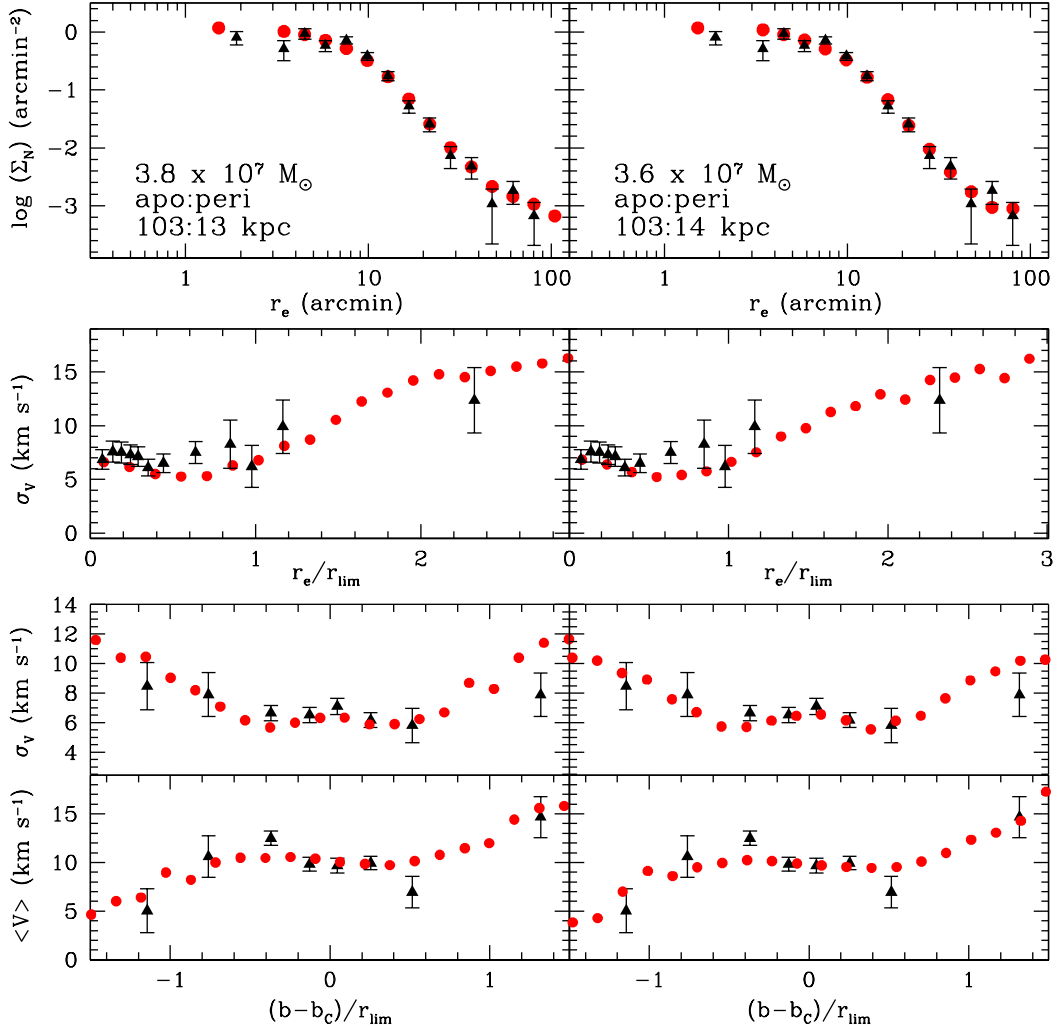}{5.50in}{0}{75.}{75.0}{-235}{-120}
\caption{From top to bottom:  Number density profiles, velocity dispersion versus
radial distance, velocity dispersion as a function of major axis, and
mean RV trend (with velocities in the Galactic Standard of Rest frame) for
our two best matching models.
Model 502I is shown by the left panels and model 502G by the right ones.
The data for Carina from \citet{Mu06a} are shown as solid triangles.}
\end{figure*}

\subsubsection{Mass-to-light ratios}

Tables 3 and 4 also give for the different models the evolution of the satellite from its initial
conditions at each apogalacticon and at the final state, with the distance, 
best fitting central King profile parameters (see Appendix A), fractional bound mass remaining, and 
central velocity dispersion given.
As may be seen, the actual King profile radii of the satellites remain relatively constant
through time.  However, the central velocity dispersion is seen to drop as the satellite
loses bound mass, closely tracking the instantaneous mass --- as 
expected if the steady state virial theorem remains a good approximation
in the face of relatively modest mass loss rates.

If we desire the $M/L$ of the satellite to remain constant in time (i.e., dark matter
and luminous matter are lost in equal proportion for MFL models), then the 
central surface brightness of the satellite must evolve scaled to $\sigma^2$.  To demonstrate
that this actually happens,
we also give for each apogalacticon time stamp in Tables 3 and 4 a measurement of the global $M/L$ as
inferred using the standard core-fitting technique (\citealt{Illingworth76}) applied with
the velocity dispersion and core radius measured at that point in the simulation:

\begin{eqnarray}
(M/L)_{\rm tot} = {{166.5 R_{c,g} \mu} \over {\beta L_{\rm tot,V}}}
\end{eqnarray}

\noindent Application of this equation requires calibration of the models by 
the actual central
surface brightness in order to ``paint" an appropriate fraction of the particles as luminous.
For the final time stamp of the models, which are intended to
represent the present Carina dSph, we simply adopt the observed luminosity of Carina
of $4.3 \times 10^{5}$ L$_{\sun}$ from \citet{Mateo1998}.  Once this is done, we
can determine the scaling from particles to luminosity by the ratio of the 
actual luminosity to the total number of particles 
in the final time step of the model.  This scaling allows us to track the {\it measured} $M/L$ 
of the satellite as determined through core-fitting backwards over time (e.g., the ninth column
in Tables 3 and 4).  
But because we are assuming MFL, we can of course independently assess what the 
{\it actual} $M/L$ of the bound  system is (which is constant by design) by the fact that we 
know what the bound mass is at the last stage of the simulation and we are assuming the satellite 
has currently the luminosity of Carina.
Tables 3 and 4 shows that the $M/L$ derived by core-fitting reasonably
matches the actual $M/L$ of the model throughout its evolution.  We have found that this
condition breaks down only when the satellite reaches near complete dissolution, as
previously reported by \citet{Kroupa1997}. The stability of the observed $M/L$ with time
is another convenient feature of our models that frees us from concerns about special
timing requirements to match models to data.  Moreover, it is not a forced condition, but one that
arises naturally in our models.

\begin{deluxetable*}{ c c r r c c c c c c}
\tabletypesize{\scriptsize}
\tablewidth{0pt}
\tablecaption{Properties of Model 502g\tablenotemark{a}}
\tablehead{\colhead{orbit} &
 \colhead{$D$ (kpc)} &
\colhead{$r_{core}$ (pc)\tablenotemark{b}} &
\colhead{$r_{lim}$ (pc)\tablenotemark{b}}  &
\colhead{$R_{tidal}$ (pc)\tablenotemark{c}}  &
\colhead{$\sigma_{v}$ (km s$^{-1}$)\tablenotemark{d}} &
\colhead{$M_{bound}$ (10$^{7}$ M$_{\sun}$)} &
\colhead{$M_{bound}/M_{0}$} &
\colhead{$(M/L)_{cf}$ ((M/L)$_{\sun}$)\tablenotemark{e}} &
\colhead{$(M/L)_{m}$ ((M/L)$_{\sun}$)\tablenotemark{f}}}
\startdata
0 & 104   & 230   & 725  &  550     &  8.73  & 3.60  & 1.00  &  38  & 45 \\
1 & 96    & 220   & 710  &  535     &  8.48  & 3.26  & 0.91  &  38  & 45 \\
2 & 97    & 220   & 760  &  505     &  7.96  & 2.77  & 0.77  &  39  & 45 \\
3 & 106   & 225   & 770  &  490     &  7.60  & 2.48  & 0.69  &  41  & 45 \\
4 & 97    & 235   & 700  &  470     &  7.22  & 2.23  & 0.62  &  43  & 45 \\
5 & 101   & 250   & 815  &  450     &  6.83  & 1.94  & 0.54  &  47  & 45 \\
Carina\tablenotemark{g} & 101 & 258 & 846 &  &  6.97 & 1.76  &  & 41 &    \\
\enddata


\end{deluxetable*}

\begin{deluxetable*}{ c c r r c c c c c c}
\tabletypesize{\scriptsize} 
\tablewidth{0pt}
\tablecaption{Properties of Model 502i\tablenotemark{a} }
\tablehead{\colhead{orbit} &
 \colhead{$D$ (kpc)} &
\colhead{$r_{core}$ (pc)\tablenotemark{b}} &
\colhead{$r_{lim}$ (pc)\tablenotemark{b}}  & 
\colhead{$R_{tidal}$ (pc)\tablenotemark{c}}  &
\colhead{$\sigma_{v}$ (km s$^{-1}$)\tablenotemark{d}} &
\colhead{$M_{bound}$ (10$^{7}$ M$_{\sun}$)} &
\colhead{$M_{bound}/M_{0}$} &
\colhead{$(M/L)_{cf}$ ((M/L)$_{\sun}$)\tablenotemark{e}} &
\colhead{$(M/L)_{m}$ ((M/L)$_{\sun}$)\tablenotemark{f}}}
\startdata
0 & 105  & 250   & 900  &   500       &  9.00  & 3.80  & 1.00  &  40  & 43 \\
1 & 97   & 230   & 740  &   485       &  8.70  & 3.42  & 0.90  &  38  & 43 \\
2 & 97   & 250   & 770  &   455       &  7.89  & 2.83  & 0.74  &  41  & 43 \\
3 & 106  & 230   & 715  &   435       &  7.59  & 2.50  & 0.66  &  39  & 43 \\
4 & 97   & 230   & 760  &   420       &  7.16  & 2.21  & 0.58  &  40  & 43 \\
5 & 101  & 235   & 850  &   395       &  6.60  & 1.85  & 0.49  &  41  & 43 \\
Carina\tablenotemark{g} & 101 & 258 & 846 &  &  6.97 & 1.76  &  & 41 &    \\
\enddata

\tablenotetext{a}{Run in an orbit with $R_{apo}/R_{peri}$ of 101/15 kpc.}
\tablenotetext{b}{Core and King radius from best fitting to bound population.}
\tablenotetext{c}{Tidal radius from equation (2) (\citealt{Oh1992}).}
\tablenotetext{d}{Central projected velocity dispersion.}
\tablenotetext{e}{Global Mass to Light Ratio from Core-Fitting technique applied to
the model's structural parameters.}
\tablenotetext{f}{Global Mass to Light Ratio from the known bound mass in the model. This value
is constant over time by design.}
\tablenotetext{g}{D, $r_{core}$ and $r_{lim}$ from \citet{Mateo1998}. $\sigma_{v}$
and $(M/L)_{cf}$ from \citealt{Mu06a}.}
\end{deluxetable*}

\subsubsection{Comparison to Proper Motion-Derived Orbit}

The orbital parameters of those models that give the best match to 
the observed Carina properties (Tables 3 and 4, Figure 10), are in very good agreement 
with the one previous attempt to determine the orbit of Carina directly. 
\citet{Piatek2003} measured the proper motion
of Carina using HST imaging and found that the dSph must currently be
near apogalacticon, and they derive the best estimate for apo- and perigalactica at, respectively,
102 and 20 kpc; this is in very good agreement with our results of 103 and 15 kpc. \citet{Piatek2003} 
also give an estimate for the orbital period of 1.4 Gyr; again, this matches well 
our derived period of 1.5 Gyr. Although their study does not find Carina in a polar 
orbit (the inclination of their derived orbit with respect to the Galactic plane is 
$39^{\circ}$), the 95\% confidence interval for their derived inclination 
is so wide $(23^{\circ},102^{\circ})$ that it includes a polar orbit such as we 
derived here\footnote{Our own derived orbital inclination (\S2) is based solely on the general 
orientation of the extended Carina population (Fig.\ 1; see also Fig.\ 3) and this estimate 
itself is likely subject to some 10$^{\circ}$ uncertainty.}. 

In summary, our ``fine tuning" procedures have allowed us to find MFL, tidal disrupting 
satellite models that quite successfully fit the prime observational properties of 
Carina: the velocity, velocity dispersion and radial density profiles.  Moreover, along 
with other properties that arise naturally, such as the close and correct tracking of the 
instantaneous central $M/L$, these model properties (1) do not require special 
conditions (e.g., particular viewing perspectives, critical evolutionary states)
to be observed, (2) have long term stability over the life of the satellite, and (3)
do not require the invocation of extra free parameters beyond initial satellite mass, scale, 
and orbital shape.

\subsubsection{Tidally Induced Ellipticity?}

Despite being a good match for most of the structural and kinematical
properties of Carina, the only shortfall of our models that we have found
is that they fail to match Carina's observed ellipticity (0.33, \citealt{Mateo1998}); rather, 
in general the models maintain bound components that are nearly spherical.
Obviously, elongation of the model satellites is observed over very large scales, when tidal
debris starts to organize into tidal tails, but not over radii where ellipticity is seen in 
the dSphs.\footnote{The orientation of the nascent tails in the simulations agrees with the 
observed instrinsic elongation of Carina, but of course the orbital poles for the simulations
were chosen assuming that the major axis of Carina lies along its orbital path (\S2).}

\begin{figure}
\plotfiddle{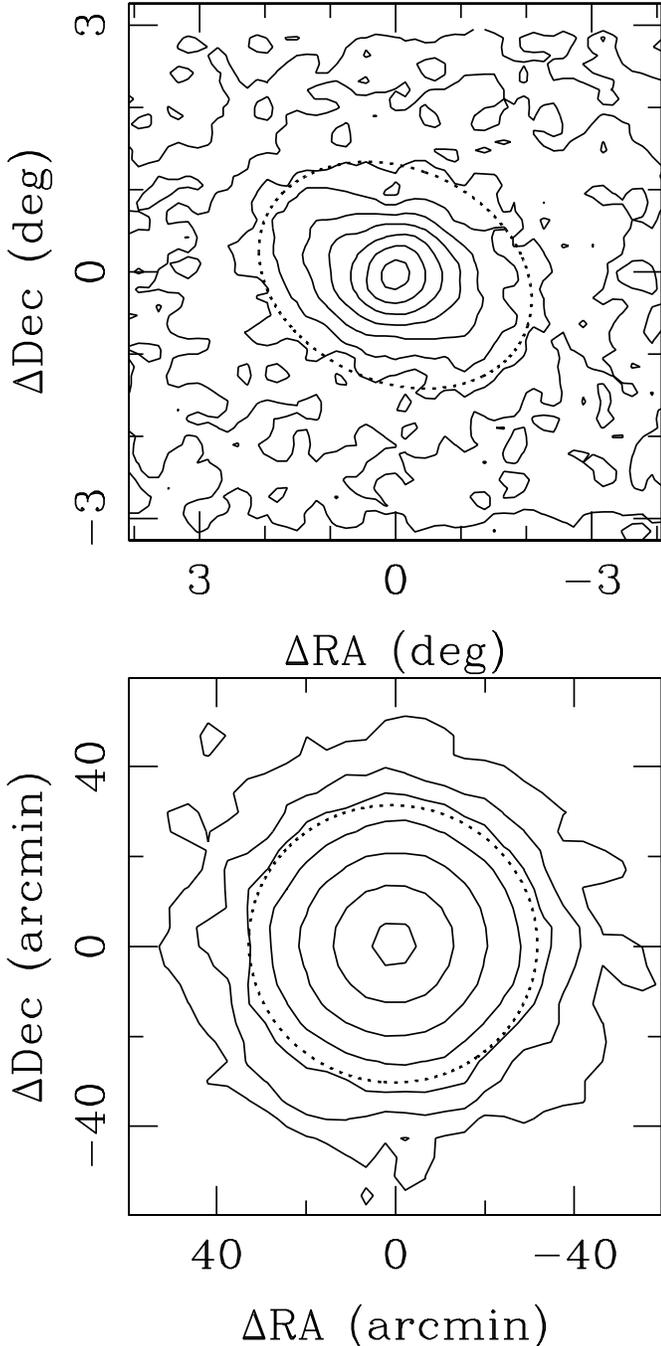}{7.00in}{0}{95.}{95.0}{-285}{-150}
\caption{{\it Upper panel}: Isodensity contours for a model satellite right before it becomes completely
unbound (5 \% of initial mass still bound). As the satellite is on the verge
of being destroyed the bound component of the satellite enlarges and elongates.
{\it Lower panel}: Isodensity contours for one of the Carina's best fitting models (502I).
Only particles within two King radii, marked by the dotted ellipse) 
have been considered to mimick the surface brightness limitations of
current photometric surveys. 
The contours have been plotted roughly evenly spaced in radius.
The density countours show a remarkably
regular structure, with no tail-like structures visible despite
the satellite losing nearly half its initial mass.
 }
\end{figure}

While spherical cores are a general result for those models that retain a significant 
fraction ($> 10$\%) of their initial mass as bound, satellites with a bound
core but having
the observed ellipticity and position angle of Carina are obtained only when
$\sim$5\% or less of the initial mass remains bound.
Figure 11 (upper panel) shows a coutour map of the density distribution for a model that is
near complete destruction.
The elongated appearance of the satellite is clear.
Interestingly, when they are near destruction, these satellites exhibit a kinematical
behavior similar to that described by \citet{Kroupa1997} --- that is, their
central velocity dispersions are significantly inflated along with an associated artificial
increase in the inferred central mass density and total mass.
However, in these cases, the satellites no longer provide a good simultaneous match to the 
density and velocity dispersion profiles of Carina, as Figure 12 shows, due, partly, to a 
sudden and significant inflation in the satellite size (by up to an order of magnitude) with 
respect to its starting scale length. We conclude, therefore, that
if our MFL models provide the correct explanation for the structure and dynamics of Carina,
the observed ellipticity in this satellite is not likely tidally induced (see \S4.2), and 
is probably related to the initial shape of the satellite.

\begin{figure}
\plotfiddle{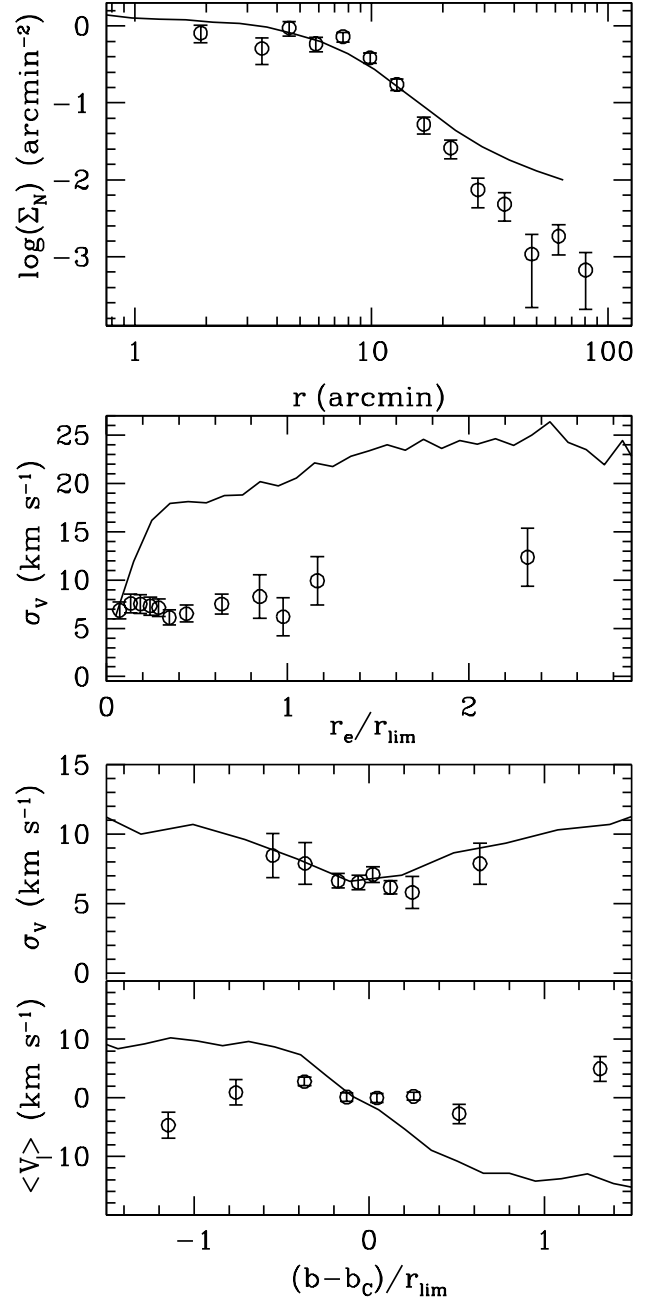}{7.00in}{0}{95.}{95.0}{-285}{-150}
\caption{Number density, velocity dispersion and mean RV trends
for the satellite shown in the upper panel of Figure 11. Data points are overplotted
for comparison as open symbols.
The model satellite has become significantly enlarged compared to its initial size,
and it has been scaled down by a factor of four to compare to the data.
While the velocity dispersion along the major axis of the model seems to provide a good match to
the data (after being scaled down), the rest of the observed properties of Carina are not
well match by the model (even when scaled down).}
\end{figure}

\section{Discussion}

Our work here has demonstrated several important features of 
tidal disruption and the dark matter content of dSphs in general, and of the Carina dSph
in particular. These results are 
at odds with some commonly held perceptions regarding
the observable effects of tidal disruption. These perceptions have partly
been driven by extrapolating the results of \citeauthor*{PP95} and \citeauthor*{OLA95}
beyond the specific cases actually explored in those papers.
We use our analysis of tidally disrupting $N$-body satellites to readdress eight statements
commonly made about dSphs and tidal disruption that we believe are either incorrect, or at least
are overly simplistic.

\subsection{Misperception 1: Lack of Rotation Signatures Means Tidal Disruption Is Not Happening}

Current RV surveys of Galactic dSphs extend now to the outskirts of these systems (as 
defined by the distribution of their luminous matter) and this provides  new constraints 
useful for checking the dynamics of tidally disrupting $N$-body simulations. \citeauthor*{PP95} 
stated that the most useful result from their $N$-body simulations is the identification of 
tidally induced rotation along the major axis of the system as a signature of tidal disruption. 
Subsequently, this has become perhaps one the most widely used criteria to assess whether 
tidal stripping is affecting particular dSph systems.

To date, kinematical surveys of most dSphs have found 
little evidence for significant rotation. This lack of a clear rotational signature
has thus lead to the conclusion that
tidal stripping of Galactic dSphs is for the most part 
unimportant (e.g., \citealt{Kleyna2001}; \citealt{Koch2007a,Koch2007b}).
However, even careful examination of the \citeauthor*{PP95} results indicates
that lack of rotation signatures within current RV surveys is not necessarily
indicative of the importance (or lack thereof) of Galactic tides. \citeauthor*{PP95}
found significant rotation along the major axis of their model satellites 
{\it only when the satellites were near
perigalacticon of their eccentric orbits and when the host galaxy's
mass distribution was modeled as a point mass}.  But PP95 also noted that when the Galactic
halo was modeled by a more realistic potential, this apparent rotation
became less pronounced in their models and was only observed over large spatial scales
(this point was also made by OLA95).
Perhaps most relevant to the typical Galactic satellites, 
\citeauthor*{PP95} saw that the rotation signal effectively disappears as the satellites 
move to apogalacticon, and this is an orbital phase near where satellites are most likely 
to be observed.  Our more comprehensive exploration of satellite parameter space and 
modeling in a more realistic MW potential also indicates that tidally induced rotation 
is not always a good discriminator of tidal disruption.
As discussed in \S3.1.2, detectable levels of induced rotation along the major axis is only 
observed over scales larger than approximately $r_{lim}$ of the model satellites, whereas 
most current RV surveys are confined to regions interior to these radii.

Thus, because tidally induced rotation can be completely unobservable
within the current spatial limits of most dSph RV surveys for satellites at typical 
points in their orbits, this test  is not reliable for {\it ruling out} tidal disruption.
Eventually, when large enough radii can be probed dynamically, this kind of test may be 
more appropiate; this is the case for Carina where RV members have been detected to at 
least 4.5$r_{lim}$ and rotation along its major axis {\it is} observed beyond its $r_{lim}$.

However, an additional complication to using this ``rotation" test over large angular 
scales is that even if a satellite is completely bound, eventually the varying projection 
of the satellite bulk motion on the radial velocity can produce an apparent ``shearing" 
trend similar to the one caused by tidal effects.  If the mean velocity as a function of 
position can be ascertained with great enough precision, it may be possible to discriminate 
the more uniform trend produced by projection from the more $S$-shaped trend 
seen in many tidal disruption models, but the differences are subtle.

\subsection{Misperception 2: Tidally Disrupting Satellites Have Pronounced Ellipticities}

An interesting result from our general survey is that for models that retain
a bound core with more than $\sim10$\% of the initial mass, 
the bound component of our satellites always 
remain spherical, with no perceptible, tidally induced elongation. 
Figure 11 (lower panel) demonstrates this situation: Here, we show the contours of the 
projected number density of one of our best matching Carina models only using particles 
within two $r_{lim}$. 
Even though some debris is projected within $r_{lim}$, the overall azimuthal 
satellite density distribution is dominated by the still bound material and, 
as is readily obvious, the contours show no substructure or clumping
and the satellite remains fairly round and regular.

This lack of tidal elongation in our model seems to be, in principle, at odds
with the findings of \citeauthor*{PP95} (see their Figure 5).  These authors found 
their simulated satellites enlarged and presented projected ellipticies
as high as 0.7, which PP95 ascribed solely to the effects of Galactic tides.
But a careful comparison between our simulations and \citeauthor*{PP95}'s indicates
that there is no actual disagreement between our models and theirs, only on the
respective interpretation of the models.
As in the case of apparent rotation (\S4.1), the tidally induced ellipticity
observed by \citeauthor*{PP95} pertains to the specific case of a
satellite being observed just before it becomes completely destroyed
and when the MW has been modeled as a point mass.
As presented in \S 3.2.4, model satellites in our survey that are near complete 
disruption, with only a small fraction of the initial mass still bound (typically 
5\% or less), also ``puff up" relative to their original size and they also typically
elongate along their major axis (Figure 11, upper panel). 

That none of our surviving model satellites present tidal elongation within their tidal 
limits suggests then that perhaps the ellipticity observed in some Galactic dSphs 
corresponds to a process other than tidal stripping, even if the satellites are currently
experiencing mass loss.  For example, if dSph morphologies are the result of tidal
stirring of disky dwarf irregular galaxies (\citealt{Mayer2001a,Mayer2001b,Mayer2007})
the triaxiality of the progenitor system may be preserved for some time.
This might also explain why, in the cases where proper motions have been measured, there is 
not always a correlation between the observed position angle of the satellites and the
direction of their orbit (e.g., \citealt{Piatek2003,Piatek2005,Piatek2006}).

Once again, our more systematic exploration of parameter space illustrates that one must be
careful not to extrapolate the results of the earlier, more specific models and expect effects
that are not necessarily general in nature or that are actually beyond the range or sensitivity 
of current surveys.  In this case, the elongation reported by PP95 and in our simulations 
is from the nascent tidal tails, and therefore, to see the effect, one must look at radii 
where the debris makes a substantial relative contribution to the bound particles (e.g., Fig.\ 13).  
This is beyond the typical surface brightness limits and/or radii of most previous 
photometric studies of dSphs (other than our own --- e.g. , 
Majewski et al. 2003, \citealt{Mu06a}, \citealt{Sohn2007}).

\begin{figure}
\plotfiddle{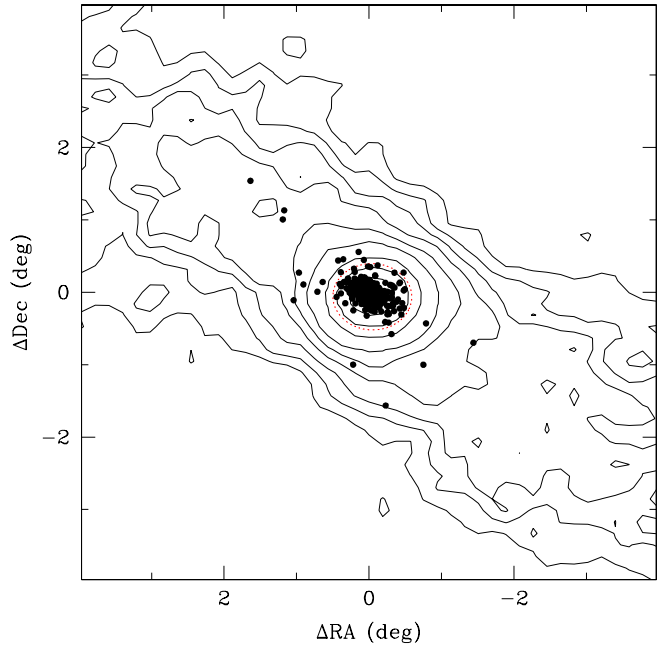}{3.30in}{0}{45.}{45.0}{-137}{-70}
\caption{Isodensity contours for the same model shown in the lower panel of Figure 11, 
but when particles over a four degree radius are considered. Again, the contours have been 
plotted roughly evenly spaced in radius along the minor axis.
The contour levels shown (from faintest to brightest up to $r_{lim}$) correspond to the following 
surface brightness levels: 35.5, 34.9, 34.6, 34.2, 33.8, 33.3,  32.0 and 30.4 mag arcsec$^{-1}$  
(assuming the central surface brightness of the model satellite is $\Sigma_{0}$=25.5 mag arcsec$^{-1}$ 
in the $V$-band). 
This time, the development of tidal tails is clearly observed,
but over scales much larger than the King limiting radius.
To illustrate how far out one needs to go to detect clear tidal tails,
current Carina RV members have been overplotted (as solid circles) for comparison.}
\end{figure}

\subsection{Misperception 3: No Detected Tidal Tails Means No Tidal Disruption}

One of the most unmistakable signatures of a system undergoing
tidal stripping is the development of debris tails which, in general,
closely trace the orbit of the satellite (e.g., \citealt{Johnston1998}).
Perhaps the most striking examples in the MW system of satellites exhibiting
tidal tails are the Sagittarius dSph (\citealt{MSWO}) and the Palomar 5 globular 
cluster (\citealt{Odenkirchen2003}; \citealt{Grillmair2006a}). Several other recently found stellar
streams have no clear progenitor but have been traced
for tens of degrees on the sky (e.g., \citealt{Grillmair2006b}; \citealt{Grillmair2006c}; 
\citealt{Belokurov2007}).
Several photometric studies have been carried out over the years to look for unmistakable 
evidence of tail-like features around other Galactic dSphs, but with the possible
exception of Carina (\citealt{Mu06a}), such evidence has been lacking.
This absence of detectable tail-like structural perturbations at large
radii, especially when combined with a lack of the (inappropriately) expected rotation signature
of \citeauthor*{PP95}, has often prompted the conclusion that Galactic tides are unimportant 
for most dSphs (e.g., \citealt{Walcher2003}; \citealt{Coleman2005}; \citealt{Segall2007}). 
However, we can use the results of our extensive $N$-body simulation survey to 
investigate the feasibility of actually
detecting tidal tails around satellites undergoing tidal disruption.

To clarify the expectations of finding tidal tails
we simulate ``observations" of our $N$-body models (where we know with certainty that
tidal tails are present). 
If one assumes that the central surface brightness
of our best matching models (502I and 502G) corresponds to the observed current value for 
Carina ($\Sigma_{0}$=25.5 mag arcsec$^{-1}$ in the $V$-band, \citealt{Mateo1998}), then
the surface brightness of their density at $r_{lim}$ would be $\Sigma \geq$ 31 mag arcmin$^{-1}$,
which is beyond the sensitivity of typical photometric methods used to explore
the surface brightness distribution of dSphs (see \citealt{PaperVI}
for detailed comparisons of these methods). Indeed, in their exploration of the formation 
of a MW-like galaxy halo by $\Lambda$-CDM infall of satellites, \citet{Bullock2005} found most 
of their resulting debris streams to be fainter than 30 mag arcsec$^{-2}$, and K. V. Johnston 
et al. (in preparation) predict there to be only one stream brighter than this on average. 
The fact that our own surveys (e.g., \citealt{PaperII}, et. seq.) can probe
these and the even fainter surface brightnesses of tidal tails is because
of the highly tuned selection for dSph giant stars, which yields much higher
signal-to-background sensitivity in the very diffuse parts of the dSph systems.
Figure 11 (lower panel) shows that ``observations" of one of our best matching Carina 
simulations (model 502I)),
to a surface brightness comparable to those of most current dSph photometric surveys
would not have detected the presence of tidal tails around the model satellite.

Figure 13 expands by a factor of four the spatial scale of the density contours
shown in the bottom panel of Figure 11 and includes
even lower surface brightness features to the point
where tidal tails are now visible (the dotted line delineates $r_{lim}$ 
for this model, a radius that is well within the regions where there are tails). 
This figure demonstrates another challenge
that photometric searches for tidal tails face:
the debris around the bound core of the dSph does not organize into
coherent tails immediately outside $r_{lim}$ but the presence of
tidal tails is only evident over physical scales much larger than this.
Therefore, much larger systematic photometric surveys to very low detection 
thresholds are needed to detect unmistakably obvious tidal tails from
real dSphs.  
On the other hand, unbound tidal debris {\it does} collect more or less radially
symmetrically surround the dSph just beyond the tidal radius, and it is the presence 
of {\it this} ``extended population" structure in the density profiles of satellites 
that is the more obvious, and more likely to be detected, expected photometric signature 
of tidal disruption.  As pointed out in Section 1.1, most of the classical MW dSph 
satellites do exhibit this very feature in their density profiles.

\begin{figure}
\plotfiddle{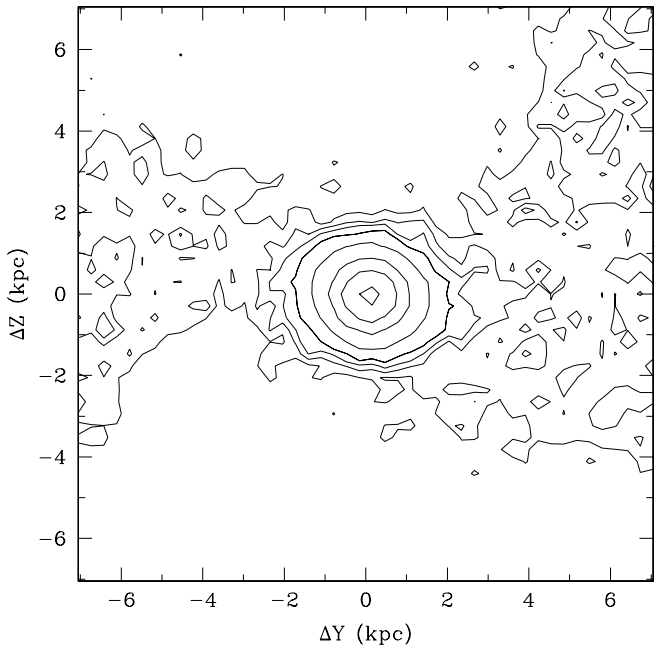}{3.33in}{0}{45.}{45.0}{-137}{-70}
\caption{Isodensity contours for the same model and particles shown in Figure 13, but
this time from a line of sight nearly perpendicular to the plane of the orbit.
As it can be seen, the expected, tidally induced $S$-shape is barely discernible.
}
\end{figure}

\subsection{Misperception 4: No $S$-shape Morphology Means no Tidal Disruption}

Another common expectation for tidally disrupting systems related to the presence 
of tidal tails is the detection of stellar density contours that become progressively
more elliptical with radius along with a variation of their position angles, which together
result in the typical $S$-shape morphology associated with tidally stripped systems 
(e.g., \citeauthor*{PP95}; \citealt{Combes1999}; \citealt{Choi2002}).  Examples of these 
morphologies have been detected in Local Group systems already proven to be disrupting,  
such as the M31 satellite NGC 205 (\citealt{Choi2002}) 
and the Galactic globular cluster
Palomar 5 (\citealt{Odenkirchen2003}).\footnote{Other systems, 
like the And I (\citealt{MI2006}) and Ursa Minor dSphs (\citealt{Palma2003}) show this 
$S$-shape morphology, but in these cases, the presence of this feature has been
adopted as one of the arguments that these dsphs are tidally disrupting.}
Thus, given the apparent connection between S-shaped morphologies and 
tidal disruption, one might be tempted to use the lack of detection of this morphology to argue
that a system has not undergone tidal interaction
with our Galaxy. 
However, closer examination of cases where this $S$-shape 
morphology has been detected shows that usually, with the clear exception of Ursa 
Minor (discussed further below), the isophotal twisting occurs over physical scales larger than $r_{lim}$.
 
In addition, previous simulations (e.g. \citealt{Combes1999}) as well as our own presented 
here demonstrate that, depending on the orientation of the satellite and 
its tails to our line of sight, this $S$-shape morphology may not be seen at all. Our best 
matching models for Carina demostrate this situation. In Figure 13, where the tidal tails of 
the satellite are clearly seen, no classical $S$-shape is observed. 
In this case, the lack of detection is readily explained by the fact that this isophotal 
twisting occurs {\it in the plane of the orbit}, and, given the Galactocentric 
distance of Carina, which like most of the classical Galactic dSphs, is large 
($>60$ kpc for most cases), the projection of the orbit on the sky 
is not favorable for observing this morphological feature (i.e., the Sun lies relatively
close to the orbital plane for distant Galactic satellites).  Moreover, even if viewed in
the correct orientation, the $S$-shape might be very subtle. Figure 14 shows the density 
contours for the same model and particles shown in Figure 13, but this time from a line 
of sight nearly perpendicular to the plane of the orbit (e.g., projected onto the Galactic 
$Y_{GC}-Z_{GC}$ plane). Even in this preferred orientation for observing the $S$-shape
it is barely visible.

The case of Ursa Minor, which shows an S-shape morphology (\citealt{Palma2003})
but where the reported $S$-shape is seen well inside 
its $r_{lim}$, is thus unique and intriguing. This dSph is at a heliocentric distance of 66 kpc 
\citep{Mateo1998}, and so it is unlikely that its orbital projection on the
sky is favorable for viewing the $S$-shape, 
regardless of the shape of the orbit.  Moreover, even 
if this were the case, we would expect to see isophotal twisting over scales larger than 
$r_{lim}$. In this particular case, the $S$-shape may be a coincidence, with the isophote 
bending the result of other physical effects at work to produce internal substructure. 
For example, the well-known secondary 
peak in the $2D$ stellar density distribution of Ursa Minor, displaced $\sim300$ pc 
from its center, and shown to be dynamically cold by \citet{Kleyna2003}, 
could potentially result in isopleths that just happen to mimic the S-shape twisting 
even if they are not of tidal origin, at least in the classical way.

\subsection{Misperception 5: Flat/Rising Velocity Dispersion Profiles Require Extended Dark Halos}

Most current RV surveys of dSph satellites yield velocity dispersion profiles
that are more or less flat or rising all the way to the $r_{lim}$ of the system.
Recently, these velocity dispersion trends have been attributed to non-MFL, multi-component 
mass distributions in dSphs, with perhaps some possible additional contribution from velocity 
anisotropy (e.g., \citealt{Lokas2002}; \citealt{Kleyna2002}; \citealt{Lokas2005}; 
\citealt{Mashchenko2005,Mashchenko2006}; \citealt{Walker2006a}; \citealt{Koch2007a,Koch2007b}). 
These multi-component models
yield both total dSph masses that are
up to an order of magnitude higher than those obtained by traditional core fitting
techniques, as well as physical scales much larger than those of the observed
light distribution, suggesting that dSphs are currently surrounded by
extended dark matter halos.  In some cases the derived satellite
total masses approach the 10$^{9}$ - 10$^{10}$
M$_{\sun}$ range (\citealt{Lokas2002}; \citealt{Mashchenko2005,Mashchenko2006};
\citealt{Read2006b}).

However, our analysis of dSph properties within the context of MFL, tidally disrupting
admits a workable alternative interpretation of the velocity dispersion profiles of dSph galaxies.
We summarize our results with respect to the behavior of the velocity dispersion profile as follows:

\begin{itemize}

\item The most important observable dynamical effect of Galactic tides is to inflate the velocity
dispersion at large radii and to produce a velocity dispersion profile that can go
from slowly declining to flat or even rising  depending on the amount of
tidal debris released and disruption history of the satellite.

\item The central velocity dispersion of systems that retain a bound core
remains unaffected by tides, which means that its value is a good reflection
of the underlying bound mass of the satellite, even when substantial tidal
stripping may be ongoing.

\item Completely flat velocity dispersion profiles are, in general, only achieved
when the satellite has become completely unbound. The size of the
velocity dispersion is driven by a complex combination of the initial satellite
mass, density, orbit and number of prior orbital passages.

\item The inflation of the velocity dispersion as a function of radial distance 
is not necessarily an effect of ``cylindrically-averaged" (i.e., azimuthally averaged) 
dispersions as suggested by \citet{Read2006a}. 
Intrinsic inflation is observed even when the velocity dispersion is studied along the major axis
(see Figures 3 and 6).

\item Completely or mostly bound systems show velocity dispersion profiles that
peak at the center and decrease with radius.

\end{itemize}

Thus, there are at least two viable explanations for the flat/rising velocity dispersion 
profiles of dSph galaxies.  However, beyond just explaining the velocity dispersion profiles,
our MFL tidal disruption models are the first models that try comprehensively to match
{\it all} presently observed general features of dSphs (except ellipticity; see \S3.2.4), 
and that satisfy all specific constraints of the presently best-studied dSph in particular.
No multi-component, tidal-stripping-free modeling of the Carina dSph has yet been 
attempted (several studies have tried to match other dSphs with more limited data sets, but Carina 
is currently the system with the most extended radial coverage, besides the Sgr dSph), so it 
remains to be seen whether such models can successfully match these observations as closely and with
as much simple elegance as tidal disruption models.\footnote{\citet{Wu2007} has created 
multicomponent dSph models that produce both flat velocity dispersion profiles and
breaks in the density profiles of several other dSphs with less radially extensive velocity data.}

\subsection{Misperception 6: Dark Matter and Tidal Disruption in dSphs are Mutually Exclusive}

Previous investigations of tidal effects in dSphs have often proceeded with an ``all or nothing" 
approach with regards to the dark matter content of these systems.  For example, \citet{Kuhn1993}, 
\citet{fleck2003}, \citeauthor*{PP95}, \citeauthor*{OLA95}, \citet{Kroupa1997} and \citet{GFDM03}, 
all explored dark-matter-less systems, while most of the references in the previous subsection 
generally sought to explain the outer properties of dSphs (most often the dynamical, but not 
necessarily the density, properties) entirely with extended dark matter halos.  Our approach has 
been to explore dark-matter-dominated systems, but ones where tidal disruption is still a prominent
feature shaping their dynamical and structural character.

Regarding the mass content and $M/L$ of our test case, the Carina dSph, 
our modeling yields revealing insights:

\begin{itemize}

\item Carinas with no dark matter (current $M/L \sim 1-3$, 
or $M\sim1\times10^{6}$ M$_{\sun}$) are hard to make:
Low mass models on benign (rather circular) orbits with scale lengths comparable
to real dSphs and that manage to survive (for 10 Gyr or five orbits)
do not show the proper density nor velocity dispersion profile. 
On the other hand, low mass models on highly radial orbits are generally destroyed 
relatively easily (as has previously been found by \citeauthor*{PP95}) and in order 
to make such satellites survive until the present time the initial densities have to 
be so high that the scale lengths are unrealistically small compared to dSphs.  
Moreover, during those evolutionary phases when the models still retain a bound core, 
the central velocity dispersion is also too low to provide an acceptable match to the 
data even if one resorts to the effects of binary stars or atmospheric jitter in the 
giant star atmospheres to account for the differences between the model and observed
central velocity dispersion (see \S1.1).

\item On the other hand, MFL Carinas with current masses significantly higher
than 3 - 4$\times 10^{7}$ M$_{\sun}$ 
are impossible to make conform to the observed velocity dispersion and density
profiles simultaneously.
They are harder to disrupt compared to lower mass
satellites for a given orbit and length scale, and even when they do disrupt, the debris
is often too energetic compared to the observations (\S2.2). In these cases, 
if one tries to 
reproduce the Carina number density profile (by varying the mass
loss rate) the models need to be put in very eccentric orbits, which, in turn, ``overheat" 
the particles and produces velocity dispersion profiles that significantly depart from that 
actually seen in Carina.
Equivalently, if one tries to match the velocity dispersion profile by increasing
the density (and therefore produce a modest mass loss rate), then the physical scale of the 
satellites becomes unrealistically small. 

\item Our best matching models (those with initial masses of $3.6-3.8\times 10^{7}$ 
M$_{\sun}$) independently predict present masses for Carina  consistent with those 
derived for the real dSph from the core-fitting method ($\sim2\times 10^{7}$ M$_{\sun}$, 
\citealt{Mateo1993}; \citealt{Mu06a}).  At first this may seem unsurprising, given that 
the initial mass and length scale of the model Carina were specifically tuned so that 
the final stage of the simulated satellite matched its observed central velocity dispersion
and core radius.
It might be argued that the agreement between the estimated mass of Carina
(based on its central velocity dispersion) and that of the simulated model is 
just a reflection of the fact that our models meet (by design) the assumptions used by 
the core-fitting technique: a single-component, isotropic, MFL, spherically-symmetric 
distribution in equilibrium follows the distribution assumed by \citet{Illingworth76} 
for estimating the total mass.

But this agreement between model and observations is not trivial.
While providing by design a good match to the observed {\it central} velocity
dispersion and scale length, the masses for the best-fitting model Carinas also
provide an excellent match to the velocity and velocity dispersion profiles at 
large radii (when represented either as a function of radial distance or position 
along the major axis) out to the limits of the current observations.  In addition, 
the amount of debris released, which is also a function of the instantaneous mass 
content, provides a very good match to the observed number density distribution.
Interestingly, the core fitting technique, applied to the model Carina, yields the 
proper instantaneous bound mass value even after the satellite has lost most of 
its initial mass, and this technique for gauging the mass only breaks down at the 
point of near total destruction of the satellite.  At all intermediate time steps 
(as Tables 3 and 4 show) the central velocity dispersion effectively tracks the 
bound mass content of the satellite. 

\end{itemize}

In the end, the ability of our best fitting model to simultaneously match all the observed 
properties of Carina (except, of course, the observed ellipticity, \S3.2.4) 
leads us to the same finding as \citeauthor*{OLA95} in that the central velocity dispersion 
of modeled dSphs generally reflects their equilibrium value even if the satellite is undergoing 
significant tidal disruption; therefore core-fitting is validated as a useful mass indicator if,
even in the presence of tidal disruption, dSphs are adequately represented by MFL models.
In the case of Carina, we show that the dSph is well described by tidally disrupting MFL 
models, even though it is still a very dark matter-dominated satellite.  Clearly, it is 
possible for dark matter-dominated models to exhibit luminous tidal disruption.

\subsection{Misperception 7: The King Limiting Radius is Meaningless as a Tidal Radius Indicator}

As discussed in \S 4.7, at present, the underlying dark matter distribution 
in dSphs is still unknown, and we are left only with the distribution of the 
luminous matter to understand the structural properties of these systems. A 
relevant question for the MFL models explored in this paper then is, if in 
fact present dSphs can be modeled well as single-component objects, 
does $r_{lim}$ provide any useful information regarding the true tidal radius 
of the satellite?

The tidal radius of a satellite can be defined as the place at which stars 
within the satellite become unbound to the system and, in turn, become bound 
to the host galaxy.  Early studies of globular clusters 
(\citealt{vonHoerner1957}; \citealt{King1962}; \citealt{Oh1992}) 
derived simple relations to estimate this boundary that depended only on the 
mass of both the satellite and the host galaxy as well as the shape of the 
satellite orbit, but soon afterward it was appreciated that the tidal radius 
of a system was also a function of the type of orbit the stars within the 
satellite have; stars on prograde orbits are more easily stripped than stars 
on radial orbits, and the latter are more easily stripped than stars in 
retrograde orbits (\citealt{TT1972}; \citealt{Keenan1975} and more 
recently \citealt{Read2006a}).  Thus, it became clear that the true tidal 
radii of dSphs are not a well defined edge, but rather a shell-like zone. 
Previous $N$-body simulation studies (e.g., \citealt{Johnston1998}; 
\citealt{Klimentowski2007}) and the simulations presented here, show that 
unbound debris can linger deep inside the space defined by the bound population 
and, at the same time, stars still bound to the satellite can remain well into 
the region dominated by tidal tails, effectively ``blurring" the ``tidal 
boundary" from a Roche-like surface to a population overlap zone.

Despite the impossibility to define a clear, well delineated tidal boundary, 
if one knows the bound mass for a satellite in a given orbit with eccentricity $e$, 
the tidal radius at perigalacticon can be estimated using (\citealt{Oh1992}),

\begin{equation}
R_{tidal} = a \left( {M_{\rm dSph}\over {M_{\rm G}}} \right)^{1/3} \left  \{ {(1-e)^2 \over {[(1+e)^2/2e]{\rm ln}[(1+e)/(1-e)]+1}} \right \}^{1/3}
\end{equation}

\noindent where $a$ is the orbital semimajor axis for the satellite, 
and $M_{\rm dSph}$ and $M_{\rm G}$ are
the mass of the dSph and the MW (inside $a$), respectively.  The question is 
how does this compare to the King limiting radius?

\citet{Burkert1997} tried to estimate the enclosed mass, $M_{\rm dSph}$,
of several dSphs (including Carina) using this equation and adopting $r_{lim}$ as 
$R_{tidal}$.
He then compared the derived $M_{\rm dSph}$ to the dynamical masses (those 
derived from core-fitting), and argued (1) that the latter masses are much 
too high compared to the former masses, or (2) reversing the problem, if 
the masses of dSphs are indeed similar to their dynamical masses, 
then $R_{tidal} >> r_{lim}$ and therefore $r_{lim}$ has no meaning as a 
tidal boundary. In the latter scenario, no extratidal stars should be 
observed in dSphs because the luminous component is effectively shielded inside 
a much larger tidal boundary.\footnote{This does not preclude dSphs from having 
extended luminous components if these form part of an extended, bound halo.}

\citeauthor*{OLA95} and \citeauthor*{PP95} also explored the relation between 
$r_{lim}$ and $R_{tidal}$ and found that a satellite 
is able to survive tidal interaction (i.e., not be completely destroyed by tides) 
for several Gyrs only if its $r_{lim}$ is smaller than twice its $R_{tidal}$.

The results from our survey provide us with a possible way
to weigh in on the question of $R_{tidal}$ versus $r_{lim}$.
In Tables 3 and 4, we show the instantaneous tidal radii of our satellites as
determined by equation (4) using the bound mass of the system at that point of 
evolution and the density of the MW at those apogalactica.  As may be seen, our 
models meet the \citeauthor*{OLA95} conditions for luminous tidal stripping 
($r_{lim} > R_{tidal}$) and survival ($2 R_{tidal} > r_{lim}$) at the same time.  
However, our results are at odds with the conclusions of \citet{Burkert1997}: 
Despite having masses of a few times 10$^{7}$ (as correctly inferred by core-fitting 
in our models),  our best fitting models have a tidal radius smaller than its 
$r_{lim}$ and do show extended, {\it unbound} populations.
This discrepant result is explained by an arithmetical mistake in the application 
of equation (4) by \citet{Burkert1997}.\footnote{While rewriting equation (4) to 
solve for $M_{\rm G}$, which involves cubing the equation, \citet{Burkert1997} did 
not cube the expression within the braces on the right-hand side where the effect 
of the orbital eccentricity $e$ is included.  The error becomes increasingly 
significant as the orbital eccentricity of a system increases, but even the value 
of the expression for the case where $e=0$ is incorrectly reported in that paper.} 

If single-component models undergoing tidal stripping are adequate enough that they 
provide a functional description to the observed properties of dSphs, then
they revalidate the value of $r_{lim}$ as at least a {\it crude} approximation
for an upper limit for the $R_{tidal}$ of dSph systems 
(i.e. $2R_{tidal} > r_{lim} > R_{tidal}$).  In this regard, \citet{MSWO} 
have already argued that in the case of Sagittarius, the only dSph clearly 
in the process of being tidally disrupted, the true tidal radius has to be well 
inside $r_{lim}$ of the light distribution.

\begin{figure}
\plotfiddle{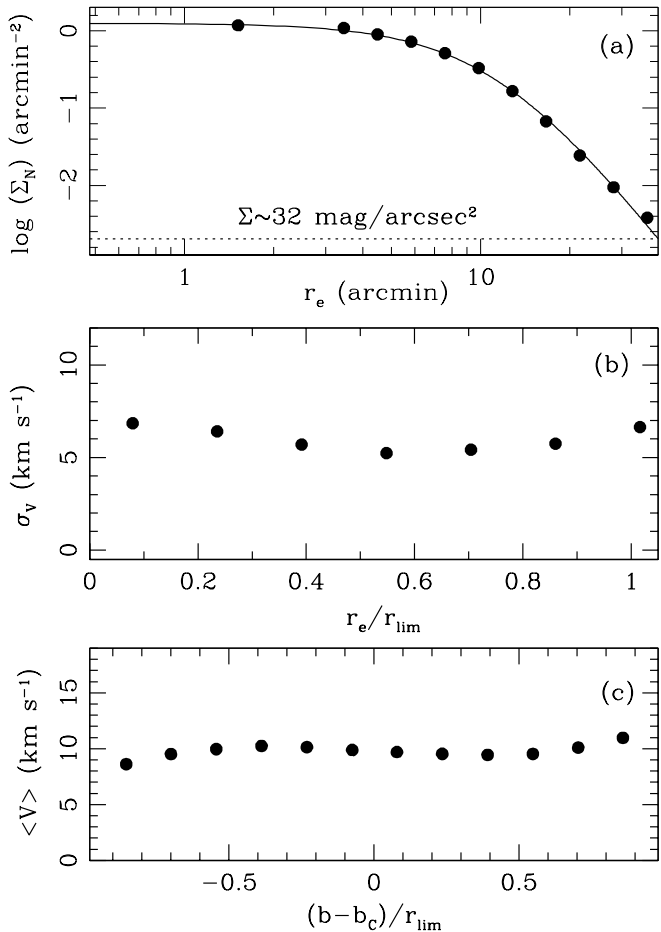}{5.10in}{0}{67.}{67.0}{-200}{-90}
\caption{(a) Number density profile, (b) velocity dispersion versus angular distance and
(c) mean RV trend along the major axis for model 502I when only particles within
$r_{lim}$ are considered. This mimics the typical surface brightness limits
of ``traditional" dSph photometric + spectroscopic studies. The dotted line in panel
(a) represents a typical background level from photometric surveys (\citealt{PaperVI}).}
\end{figure}

\subsection{Misperception 8: The Combined Observations of dSphs Show That They 
Are Not Being Tidally Disrupted}

The main purpose of this work has been to demostrate that Galactic tidal effects 
can produce the observed properties of dSphs, in particular, those of Carina, the
Galactic dSph with the most extensive structural and kinematical data sampling to 
date, but otherwise a system not unusual compared to most other dSphs in its
structure and dynamics.  We have also previously reported that the Leo I dSph is 
well-fitted by tidal disruption models (\citealt{Sohn2007}), and possesses 
characteristics (an asymmetric velocity distribution at large radius) that
cannot be readily explained by an extended dark halo. Recently, \citet{Mateo2007}
reported a rotational signal in the outer parts of Leo I they interpret as the
signature of tidal stripping, in line with our arguments for Leo I.  
While the Sgr dSph is an uncontroversial case of a dSph in the process of being 
accreted, there is a tendency to attribute Sgr (\citealt{Mateo1998}, 
\citealt{Gilmore2004}), and now Carina as well (\citealt{Gilmore2007}), as
``exceptions to the rule" that dSphs are in general not disrupting.
The results presented here suggest that it is not straightforward to argue against 
this position with current data on dSphs other than Sgr because, even if dSphs are 
undergoing significant tidal stripping, the effects can be for the most part 
undetectable (e.g., \S4.3).

To illustrate how the lack of sufficiently sensitive data can devolve
into an apparently definitive argument against tidal disruption we use recent
discussions regarding the state of the Draco dSph galaxy. 
The lack of concordance between observational results and
expectations from previous $N$-body simulations have been used to conclude 
that Draco cannot be significantly tidally influenced by the MW.
The main arguments supporting this view are summarized by \citet{Kleyna2001}, who
argue that Draco cannot be significantly affected by Galactic tides because:
(1) no significant rotation is observed within the radial extent of the current
kinematical surveys (which are limited to Draco's $r_{lim}$), (2) Draco shows a 
``roundish" appearance (low eccentricity of $\sim0.3$) in apparent contradiction 
with the predictions of the $N$-body simulations done by \citeauthor*{PP95} and 
\citeauthor*{OLA95}, and (3) its surface brightness profile is adequately fitted 
by a Plummer law without the need to resort to an ``extended" population in the 
outer parts.\footnote{More recently, \citet{W04} found that the surface brightness 
profile of Draco seems to show a ``break" at about 25', which is inside 
its observed $r_{lim}$, whereas \citet{Segall2007} could not confirm the presence of this 
extended population down to three orders of magnitude below the central number 
density.} Furthermore, \citet{Odenkirchen2001} add that no tidal tails from Draco 
are visible within the SDSS data and  \citet{Read2006b} argue that tidal stripping 
of dSphs should cause the velocity dispersion profile to rise rather than to remain 
flat, which is the observed behavior for Draco.

While not claiming by any means that Draco is being tidally disrupted, we have shown 
in this paper that an object with Draco's observed structure and kinematics can still 
be experiencing tidal stripping of its luminous matter.  To illustrate further how this 
fact may be disguised, Figure 15 shows the density profile, velocity dispersion profile 
and RV trend (in GSR) of one of our best matching model Carinas, but with data ``observed" 
only out to $r_{lim}$, as is the case for Draco (compare to Figure 10, where the full 
model data are shown).\footnote{The kinematical data on Draco in fact barely reaches 
its $r_{lim}$.} Carina and Draco are at nearly similar distances (80 vs 100 kpc),
have similar sizes, similar luminosities and both have high central $M/L$.  However 
the two dSphs have different $v_{GSR}$, with the velocity suggesting a vastly larger 
apogalacticon for Draco; nevertheless our Carina model is useful for this illustrative 
exercise.

In Figure 15a we show that, as observed in the case of Draco, a Plummer model does an 
excellent job fitting the photometry, even though, in fact, the outer parts of the model 
density profile are dominated by debris and the bound component has departed from the 
initial Plummer distribution.  Figure 15b shows that the velocity dispersion profile does 
not significantly increase beyond the central value within the radial extent of our 
``observations", a result that contradicts the above \citet{Read2006b} claim for the
expected rising behavior of the dispersion profile used to argue 
against tidal disruption in Draco.  Figure 15c shows that no statistically significant 
rotation is observed within the nominally probed Draco radius, where significant means 
greater than the velocity dispersion (\citealt{Pryor96}).
Finally, the intrinsic eccentricity of the bound component of the model, as ``observed" 
in Figure 11, is close to zero.  Both of these latter findings are also contrary to 
arguments by Kleyna et al. (2001) against Draco disruption, arguments based on 
expectations from the models developed by \citeauthor*{OLA95} and \citeauthor*{PP95}.

Taken together, the Figure 15 ``observations" of a tidally disrupting (fractional mass 
loss rate of 0.1 per Gyr), MFL satellite model --- when coupled with common misperceptions 
regarding the expected characteristics of such structures --- can be used to argue 
compellingly that our model satellites are neither tidally disrupting nor well-expained 
by MFL models. With this apparent ``failure of the tidal model" (\citealt{Klessen2003}), 
it is understandable how searches for an alternative mechanism to explain the flat Draco 
velocity dispersion profile would be deemed unnecessary or irrelevant.

\section{Final Remarks}

\subsection{What Other Studies Are Saying}

As mentioned in the Introduction, the \citeauthor*{OLA95} and \citeauthor*{PP95} 
studies provided important benchmarks in the exploration of the structure and dynamics 
of dSph galaxies, but in many ways the results of these studies have been taken out of 
context and extrapolated to regimes that, we believe, were not actually intended by 
these authors.  {\it The primary emphasis of these two studies was to establish whether 
tides could inflate the central velocity dispersions of dSph galaxies as a way to 
eliminate the necessity of dark matter; the goal of these studies was not to rule 
out that \underline{any} tidal effects could be occurring in dSph satellites.} 
Even though \citeauthor*{PP95} and \citeauthor*{OLA95} succeeded in demonstrating that 
in fact tidal effects {\it don't} typically inflate central velocity dispersions (a result 
we have also found here), it doesn't follow that tidal disruption is not happening in 
dSph systems.  Nor is it appropriate to take the observed effects in the very well-defined
and restricted range of the \citeauthor*{PP95} and \citeauthor*{OLA95} simulations 
and apply them generally to other contexts.

For example, perhaps the most cited broadside used to 
sink the tidal disruption proposition is the lack of observed dSph rotation, 
--- e.g., from \citet{Koch2007b} : ``... the most efficient way of gathering evidence whether a 
particular dSph is being affected by tides sufficiently strongly that its kinematics are 
being modified is to look for any sign of apparent rotation in its outer regions."
This leads to arguments against tides not only in particular dSphs --- e.g., ``all in all, 
there is no kinematical evidence of any significant velocity gradients in Leo II, either 
due to rotational support or produced by Galactic tides. It therefore appears that tides most 
likely have not affected Leo II to any significant degree..." (\citealt{Koch2007b}) --- 
but against tides generally in Local Group dwarf galaxies, since ``...to date, 
all but one of the dSphs of the Local Group show no significant velocity gradients."
But our analysis of this manifestation of tidal disruption (\S 3.1.2 and \S4.1) shows 
clearly that rotation is not a reliable test of tidal disruption --- a result that, in 
hindsight, should not be that surprising since (1) \citeauthor*{PP95} {\it already had 
demonstrated that the effect was pronounced in only a limited set of circumstances} 
(e.g., at perigalacticon, or when the MW is modeled as a point mass), (2) both 
\citeauthor*{PP95} and \citeauthor*{OLA95} stress that this phenomenon would be visible 
only at very large radii (i.e., outside the tidal radius, typically on $>1$ deg scales) 
where few dSphs have been explored, and (3) with small amplitude in most cases.

Meanwhile, in addition to the rotation argument, \citet{Gilmore2007} suggests that
{\it flat dispersion profiles also argue against tidal disruption}: ``...there is some evidence
... for flat velocity dispersion profiles to large distances (\citealt{Sohn2007}; \citealt{Mu05}), ... 
which are inconsistent 
with simple tidal disruption effects, particularly since most dSphs show
no evidence of apparent rotation (\citealt{Koch2007a})."  
But of course, in addition to our modeling here,
tidal disruption models have already been demonstrated to be capable of 
producing flat dispersion profiles in a number of studies in preceding years 
(\citeauthor*{PP95}; \citealt{Kroupa1997}; \citealt{Mayer2002}) ---
including in the very Sohn et al. study referenced in the above passage.  Indeed, the dominating influence
of unbound stars on the velocity dispersion at large radius is one of the very reasons MFL
tidal disruption models have been invoked (at least by our group --- e.g., \citealt{Law2005};
\citealt{Sohn2007}) as a potential dSph scenario.  In actuality, it is {\it only} in this circumstance
that such models {\it could} work, for, as summarized by
\citet{Gilmore2007}:
``More generally, there are unavoidable consistency requirements in any mass follows
light model. In any model where mass follows light the projected velocity dispersion must
be maximum at the centre, and then fall monotonically. For a well-mixed (star cluster)
system, the velocity dispersion will decrease by roughly a factor of two over three core radii.
This is an unavoidable requirement for any mass follows light system, and is observed in star
clusters ... Such a velocity dispersion profile is not required by data in any well studied
galaxy, however small, further emphasising the intrinsic difference between (virial,
King) star clusters and galaxies."  The latter description as written is obviously correct 
{\it in the absence of tidal effects}, but it should not be used as an argument against MFL models 
when they are being used to explore the very regime --- tidal disruption ---
to which the statement is {\it not} applicable.  

Perhaps another unfortunate aftermath to the \citeauthor*{PP95} and \citeauthor*{OLA95} papers 
is that, by their example of exploring {\it completely dark-matter-free} satellites, these 
two studies may have had an unintended and undue influence in paralyzing subsequent 
discussions into considering primarily two polarized viewpoints or modeling 
strategies --- namely, those where (1) the model satellites have {\it no} dark matter 
(thus requiring some reassessment or proposed revision of the physics involved with 
conventional mass determinations to reconcile the DM-less hypotheses with the 
observations), or where (2) dark matter is invoked to be responsible for {\it all} 
observed effects seen in dSph galaxies (most starkly demonstrated by the frequent 
incorporation of large dark matter halos to explain the structure and dynamics of dSphs 
at large radii).  In our opinion, and as demonstrated by the results presented in this 
paper, there is an enormous middle ground between these polar positions.

Nevertheless, more recently the weight of discussions has strongly tacked to the side 
of the large dark matter halo hypothesis to explain the structure of dSphs --- to the 
point that there are often strongly expressed sentiments that there simply is no longer 
any reason for consideration of alternative models.  Ironically, this viewpoint 
is partly driven by, on the one hand, the apparent success of $\Lambda$-CDM in explaining 
structure on large scales, but on the other hand the {\it lack of success} of $\Lambda$-CDM 
in explaining structure on galaxy scales so that more massive satellites are needed to 
help resolve the missing satellites problem.  Extended, non-MFL dark matter halos can not only account
for some of the observed properties of dSph galaxies at large radii and fit into
$\Lambda$-CDM paradigm, but they mitigate to some degree the mismatch in the observed and
$\Lambda$-CDM-predicted mass spectrum of satellite systems.
Thus, for example \citet{Wu2007} concludes ``Because our [multi-component, 
extended dark matter] models can fit both radial velocity profiles and surface number 
density profiles, the so-called `extra-tidal extensions' in the surface number density 
profiles found by \citet{IH95} and \citet{W04} do not require any special ad hoc 
explanation. {\it Thus it is not valid to consider them as the evidence of tidal 
stripping}, as proposed by some authors (e.g., \citealt{Delgado2001}; \citealt{GFDM03}; 
\citealt{Mu05})."\footnote{Emphasis added.}  

But we have shown here that our tidally disrupting MFL models can fit {\it not only} 
the radial velocity profile and surface number density profile of the Carina 
dSph {\it but also} the velocity dispersion profile. 
In addition, tidal disruption is a naturally occurring, established phenomenon 
in the evolution of galaxies (see \S5.2).  Thus, the MFL models are less ad hoc than implied 
above and, indeed, they provide a fully self-consistent explanation for all of the structure 
and dynamics of at least the Carina dSph.  Unfortunately, what such models are {\it not} 
consistent with is the prevailing understanding of dark matter, wherein the latter is non-baryonic 
and does not strongly couple with luminous matter.  Thus it is not clear {\it why} mass should 
follow light.

On the other hand, while multi-component models are aligned with the $\Lambda$-CDM orthodoxy, 
there are, as yet, no specific predictions in the context of dSphs from $\Lambda$-CDM theory 
for the expected structure and dynamics of the luminous matter embedded in dark halos. 
Indeed, the inability to predict both the observed cored profiles and the angular momentum 
distributions of galaxies are acknowledged additional failings of CDM on galactic 
scales (e.g., \citealt{Navarro2000a,Navarro2000b}), 
and multi-component models of dSphs {\it must also} adopt empirically-based, ad hoc 
distributions for the luminous matter, such as King profiles (e.g., \citealt{Strigari2007}; 
\citealt{Penarrubia2007}).

Clearly then each the MFL and multi-component models have unsatisfying 
philosophical aspects and neither can account for {\it why} luminous matter should
organize in a particular way in the presence of substantial dark matter; under such 
circumstances we believe it is pertinent to address more basic questions such as
``Which model works better?" or ``Can a given model be definitively ruled out?".  
One of the goals of this paper is to try to ``clear the air" with regard to (MFL) tidal disruption models, 
i.e. to clarify their expected characteristics over a wider (and more likely) range of parameter space
than previously explored, 
so that, ultimately, stronger tests can be developed for either verifying the tidal disruption
hypothesis or ruling it out.
Ironically, while we can make testable predictions for the tidal disruption hypothesis that can be 
checked with the discovery and observations of dSph stars at ever large radii, 
there is no direct way to check extended
dark matter models at large dSph radii, beyond where there are sufficient luminous tracers and
where most of the dark matter halo is supposed to lie.  
In this sense the extended dark matter models are virtually unconstrained, 
because the extent and mass of the dark matter halos depends strongly on how one constructs the 
mass function (or $v_{circ}$ distribution function) beyond the light.  And, even at radii where 
some stars are present, with extra free parameters at their disposal, multi-component models can 
usually be constructed to accommodate the properties of those stars.
Perhaps this is why tidal disruption models are ``more vulnerable" in the common
discourse on dwarf galaxies: Whereas the challenging exercise of
discovering dSph-associated stars at large radii is the {\it sine qua non} of the tidal disruption scenario, 
extended dark matter halo theories are essentially immune
to whether stars are found at large radius or not.  
At this point it seems that virtually the only way to {\it disprove} the existence of 
an enormously extended dark matter halo around any particular dSph is to push 
the discovery of associated stars to {\it reductio ad absurdum} radial limits where 
unreasonable dSph masses are needed to keep the stars bound to the satellite (a strategy 
employed in, e.g., \citealt{Mu05,Mu06a}).

\subsection{What We Are Saying and What We Are {\it Not} Saying}

In this paper we attempted to go beyond the seminal \citeauthor*{PP95} 
and \citeauthor*{OLA95} papers to explore tidal disruption models across a 
wider, and we believe more realistic, part of the parameter space of satellite 
mass, density, and orbit.  Like these landmark papers, we used one component 
models, but with a key difference in that we adopt satellite masses consistent 
with dSphs having significant dark matter contents --- i.e. consistent with the 
central mass-to-light ratios.  

Remarkably, despite the simplicity of our models, which by virtue of having a 
single dynamical component invoke the minimum number of free parameters (satellite 
mass, linear scale, and orbit), we have been able to zero in on examples that 
rather faithfully reproduce Carina's velocity profile, velocity dispersion profile, 
density distribution and other properties over its entire sampled radius, and without 
appealing to special timing or other particular viewing conditions (except that the 
satellites match the current properties of the real Carina system).  We have previously 
also successfully modeled the properties of the Leo I dSph system using similar 
techniques (\citealt{Sohn2007}).\footnote{We have used these techniques to model 
the Sgr dSph system (\citealt{Law2005}) as well, but in this case the focus was less 
on matching the properties of the parent satellite core and more on adequately explaining 
the properties of the tidal debris.}  In both cases, the defining characteristic of the 
successful family of models is high orbital eccentricity.  And in both cases the central 
velocity dispersion is found to faithfully represent the bound mass and central $M/L$ 
of the parent satellite (40 and 5, respectively, for Carina and Leo I), while the 
properties of the outer part of the satellite (the density law, velocity profile, and 
velocity dispersion profile) are strongly influenced by the presence of unbound tidal 
debris.  But, far from being {\it ad hoc}, as characterized and dismissed by \citet{Wu2007}, 
this tidal debris is a {\it natural consequence} of the dynamical evolution of the 
satellite and, moroever, {\it it represents a phenomenon already observed in Nature}:
Indeed, were one to subscribe to 
the Principle of Parsimony, a successful one component model would be {\it preferred}
over an equally successful multi-component model, which could be argued to contain
superfluous elements\footnote{One of the many variants of the philosophical maxim often 
called Occam's Razor is by Thomas Aquinas: ``If a thing can be done adequately by means 
of one, it is superfluous to do it by means of several; for we observe that nature does 
not employ two instruments where one suffices." ({\it Summa Contra Gentiles}, from 
the {\it Basic Writings of St. Thomas Aquinas}, translated by A.C. Pegis, Random House, 
New York, 1945, p. 129).}.

Our own position regarding the implications of the simulations presented in this 
paper takes into account not only the practical success of tidally disrupting, MFL 
models in describing at least some specific MW dSph systems, but several additional, 
more general observations:

\begin{enumerate}

\item Both dark matter and tides exist in the Universe.  Tidal effects are expected in 
CDM models (e.g., \citealt{Hayashi2003}; \citealt{Kazantzidis2004}; \citealt{Kravtsov2004}; 
\citealt{Bullock2005}; \citealt{Knebe2006}).

\item The degree of importance of tides must surely vary among Galactic satellites, 
being a function of the structure of the satellites and the nature of their orbits.

\item We see tidal debris streams throughout the Galactic halo, some of them massive 
or dynamically warm enough to have come from the disruption of dwarf galaxies 
(\citealt{Newberg2002}; \citealt{Majewski2004}; \citealt{Duffau2006}; 
\citealt{Grillmair2006b,Grillmair2006c}; \citealt{Belokurov2007}).

\item There are growing suggestions of apparent tidal effects among MW dSphs besides
Sagittarius, Carina and Leo I, including most recently in the Ursa Major 
\citep{Willman2005,Willman2006}, Ursa Major II \citep{Zucker2006b,Fellhauer2007}, 
Canes Venatici \citep{Zucker2006a}, Bo\"{o}tes \citep{Belokurov2006}, and 
Hercules \citep{Coleman2007} systems.

\end{enumerate}

\noindent A theory for the structure and evolution of dSph galaxies that can also 
account for the above, generally non-controversial elements in a natural way would 
have the appeal of an overall economy of hypothesis.

While we have not ruled out large dark matter models and our $N$-body simulations 
are by no means intended to be unique descriptions of the Carina dSph in particular, 
or dSphs in general, we hypothesize that at least {\it some} present dSphs might 
actually be well represented by tidally disrupting, MFL models.  This hypothesis is 
not incompatible with $\Lambda$-CDM, and it is not our intention to argue against 
$\Lambda$-CDM or extended dark matter halos in general.  It would be foolish to do 
so, for
it is virtually impossible to prove that extended dark halos do not exist, whereas, 
as we have discussed in \S4, it is {\it very} difficult to prove that tidal disruption 
{\it is} happening in most dSphs.  Under these circumstances, the most conservative 
position is to say that both possibilities might coexist.

For example, if most luminous galaxies start out embedded in extended dark subhalos 
of various relative sizes, eventually those on more disruptive orbits and/or with lower 
densities and/or radial extents will have their dark matter cloaks whittled down by tidal 
mass loss until their luminous matter is ``exposed" and vulnerable to the same tidal 
effects (creating visible tidal debris streams).  Indeed, \citet{Klimentowski2007} show 
that cosmologically-driven galaxy formation models with two-component satellites
that also experience tidal disruption naturally lead to the evolution of MFL-like satellites 
after their extended dark matter halos have been stripped down to the luminous cores, and
at the present time a certain fraction of Galactic subhalos can be expected to be in this
state.  Thus, {\it the Carina dSph didn't necessarily start out as an MFL-like satellite,} 
even if it might behave as one now.  Other dSphs, such as Draco, might still be firmly 
enshrouded within extended dark matter cocoons (\citealt{Lokas2002}; \citealt{Kleyna2002}; 
\citealt{Mashchenko2005}; \citealt{Segall2007}; \citealt{Wu2007}), although, again, as 
pointed out in \S4.9 and Figure 15, were Draco in a tidally disrupting phase it could be 
extremely difficult to prove.  dSphs like Ursa Minor (\citealt{Delgado2001}; 
\citealt{Palma2003}; \citealt{Mu05}) and Leo I (\citealt{Sohn2007}) might be in states 
similar to Carina as we have modeled it here, whereas other systems, such as Sgr and perhaps 
some of the newly found, extremely low luminosity dSphs with evidence of tidal disruption
(e.g., Ursa Major I and II, Canes Venatici, Bo\"{o}tes, Hercules) may be farther along this 
path of metamorphosis.  \citet{Fellhauer2007} have already shown explicitly that Ursa Major II 
--- the apparent progenitor of the ``Orphan" stream (\citealt{Belokurov2007}) --- is better 
described by a single-component than a two-component structural model, in keeping with the 
general scenario outlined here.
Finally, we might expect there to be some systems now completely destroyed, and visible only
by their tidal debris.  Such debris might be in well-correlated tidal streams, or more 
diffuse ``clouds", depending on the details of the satellite-MW interaction.  Indeed, we 
have produced some $N$-body models for systems that are of low mass, low density satellite 
on highly radial orbits that are completely destroyed, and the remains of these systems 
resemble the large, diffuse, halo ``star cloud" recently discovered in Triangulum-Andromeda 
by \citet{Rocha-Pinto2004} and \citet{Majewski2004}.  
Other features, like the ``shelves" of stars see around the disk of M31, can be created as 
tidal shells by very intense satellite-host interactions (\citealt{Fardal2006}; 
\citealt{Gilbert2007}).

Whether such a unified dSph evolutionary scenario might conflict with standard CDM 
models can be tested by comparing the predicted versus actual number of systems in each 
of the above evolutionary states. As shown by S. R. Majewski et al. (in preparation), tidal 
disruption and $\Lambda$-CDM are only {\it forced} into confrontation should it be found 
that {\it too many} MW dSphs can be well represented by MFL, tidally disrupting systems.  

We thank an anonynous referee for her/his useful comments that helped improve this paper.
RRM and SRM acknowledge funding by NSF grant AST-0307851, NASA/JPL contract 1228235
for the SIM PlanetQuest Key Project {\it Taking Measure of the Milky Way}, 
the David and Lucile Packard Foundation, and the generous support of Frank Levinson
through the Celerity Foundation.  KVJ acknowledges support from an NSF CAREER
Award, AST-0133617 as well as support from the same NASA/JPL contract above.

\begin{appendix}

\section{Some Comments on the Use of King Profiles}

Although not particularly pertinent to the results of this paper, the topic of King profiles
has often been linked to studies of the luminous structure of dSph galaxies.  In the spirit
of the overall review of dSph tidal disruption models we have presented, a few comments about
King profiles seem appropriate.
 
Historically, King profiles (\citealt{King1962,King1966}) have frequently 
been used to represent the stellar density distributions of dSphs 
(e.g., \citealt{Hodge1966}; \citealt{Eskridge1988a,Eskridge1988b}; \citealt{IH95}).
In this regard it is important to distinguish between \citet{King1962} {\it profiles},
which are a simple empirical description of the distribution of light in a 
spheroidally-shaped system, and \citet{King1966} {\it models}, which are
physically-motivated and derived explicitly for the case of densely
populated globular clusters placed in an external gravitational field.
\citet{King1966} showed that under these circumstances, his single-component
models evolved naturally to the empirical profile he derived earlier for globular
clusters, one whose structure can be described in terms of two linear physical
scales: a core and a truncation radius (identified by \citealt{King1962} as 
the {\it tidal} radius).  In globular clusters two-body stellar relaxation 
dominates the internal dynamics of the system and leads to the King profile in
an external field, but two-body relaxation is much too inefficient in low density dSphs.
Nevertheless, King profiles have frequently been found to provide an overall adequate 
{\it empirical parameterization} of the structural properties of (the most 
luminous, inner parts of) the light distribution in dSphs. In this usage, the outer
physical scale is associated with the asymptotal radius at which the fitted two-parameter
profile approaches zero (or would approach zero, in the absence of the ``break population"; 
we have identified this {\it empirically-defined} radius as $r_{\lim}$, the limiting radius, because
its connection to a true tidal radius is undefined (and, indeed, one of the controversial
questions about dSph structure --- e.g., see \S4.7).

In this paper we have used King {\it profiles} only in this way --- as a handy
empirical descriptor simply to define a {\it scale} for comparing the sizes of
the bound central cores of our simulated satellites to each other and to the 
luminous inner part of Carina itself.  In this case, a King profile is a logical
expediency motivated by the fact that our initial particle configurations are  
Plummer models (i.e., non-tidally-truncated), and as they become shaped by the
tidal field they come to resemble King profiles.  {\it But none of the conclusions in this
paper depend on the use of a King profile or any assumptions regarding its validity.}
With the growing popularity of multi-component mass models for dSph systems,
simple King {\it models} --- designed for dense, tidally truncated, single-component
systems --- would naturally be considered inadequate, and, as a result, the
King {\it profile} description  has been increasingly resisted and criticized when
applied to dSphs (e.g.,  \citealt{Kleyna2001}; \citealt{Gilmore2006, Gilmore2007};
\citealt{Koch2007a}; \citealt{Wu2007}).
Obviously, in these circumstances King models are physically meaningless:
If Galactic dSphs are not single-component bodies, but rather more complex
systems where the dark matter is distributed over much larger physical scales than
the light, then the 
physical meaning of $r_{lim}$ as a tidal truncation radius in the corresponding
King profile is lost --- and consequently, so too is the interpretation that ``breaks"
in the density distributions indicate the presence of tidal debris.

Koch et al.'s (2007a) discussion demonstrates various of the issues at play:
``It has frequently been suggested in the literature that the detection of member
stars well beyond the formal tidal limit radius of dSphs, canonically defined via
a single-component King model representation of the observed surface brightness
profiles, is evidence for physical tides. A large number of dSphs have been claimed
to show hints of extratidal stars and thus tidal disruption (Irwin \& Hatzidimitriou 1995;
Mart\'inez-Delgado et al. 2001; Palma et al. 2003; Mu\~noz et al. 2005, 2006a, 2006b).
Evidence for this has also been reported for Leo I (S06).  What such studies show,
in fact, is that the presumed parametric fit to the measured surface brightness
(and/or direct star counts) is inappropriate at large radii. Given that there is
no astrophysical basis to the application of a King model to a galaxy, an
astrophysical interpretation of the corresponding labels as physical processes
is fraught with peril."

Unfortunately, the above description
oversimplifies the arguments
made to argue in favor of the tidal disruption hypothesis by at least some of
the papers they quote.
For example, in the \citet{Mu05,Mu06a,Mu06b} papers cited above (and elsewhere), additional
information --- such as the dynamics of the stars, their two-dimensional
sky distribution,  the implied dark matter masses needed to keep the stars bound,
and the fact that the radial density profiles resemble models of disrupting
dSph galaxies --- were also weighed into those analyses and provide more powerful and
more important support for the notion that these systems may be tidally disrupting.
In the end, while the observed ``breaks" in the density distributions of dSphs
are not an unequivocal indicator of tidal stripping, they are at least consistent
with the presence of tidal debris in the outer parts of these satellites as
observed in numerous $N$-body simulations, even when multi-component models are used
for the dSphs
(e.g., \citealt{Read2006b}; \citealt{Klimentowski2007}; \citealt{Penarrubia2007}).

Finally, it is worth pointing out that there might be circumstances where King
profiles (or other two-parameter density law) are actually
{\it physically motivated} and therefore are an appropriate {\it physical}
descriptor of dSphs.  \citet{King1962} himself addressed this issue by pointing
out that his empirically derived density law is applicable to relaxed systems
{\it regardless of the relaxation mechanism}.  For example, {\it violent
relaxation} might provide an alternative to two-body relaxation, perhaps as a
process during the early formative stages of dSphs.  Alternatively, as we have
explored in this paper, dSphs might indeed be tidally truncated objects where
their flat/rising velocity dispersion profiles are the result of tidal stripping,
and thus with a truncation radius evolving in the luminous matter naturally.
Such a situation may be entirely consistent with $\Lambda$-CDM expectations
that these systems are (initially) embedded in larger dark matter halos, if the
systems evolve to where they are reasonably described by single component models
and King profiles only in the late stages of tidal evolution, after the dark outer
layers of these subhalos have already been stripped away\footnote{It might be argued
that, even after significant tidal stripping, the dark matter cocoons surrounding the
dSphs' stellar components should keep the initial, expected Navarro, Frenk,
\& White (NFW, 1997) density distribution.
However, indirect observations show that the dark matter distribution in dSphs is more
likely to be cored than cuspy (\citealt{Kleyna2003}; \citealt{Goerdt2006};
\citealt{Gilmore2007}).} (see \S5.2).  
Along these lines, \citet{Klimentowski2007} have shown that a two-component satellite 
with an initially more extended dark matter distribution can naturally evolve through 
tidal stripping into a system with a radial constant $M/L$.

\end{appendix}

\end{document}